\documentclass{iopart}

\expandafter\let\csname equation*\endcsname\relax
\expandafter\let\csname endequation*\endcsname\relax

\usepackage[acronym,shortcuts,sanitize=none]{glossaries}
\usepackage[colorlinks,linkcolor={blue},citecolor={blue},urlcolor={red}]{hyperref}
\usepackage{iopams}
\usepackage{subfig}
\usepackage[margin=10pt,font=small,format=hang]{caption}
\usepackage{multicol}
\usepackage{multirow}
\usepackage{array}
\usepackage{graphicx}
\usepackage{xcolor}
\usepackage{cleveref}

\DeclareGraphicsExtensions{%
    .pdf,.PDF,%
    .png,.PNG,%
    .jpg,.mps,.jpeg,.jbig2,.jb2,.JPG,.JPEG,.JBIG2,.JB2}

\usepackage{xspace}
\newacronym{LIGO}{LIGO}{Laser Interferometer Gravitational-wave Observatory}
\newacronym{aLIGO}{aLIGO}{Advanced \protect\ac{LIGO}}
\newacronym{iLIGO}{iLIGO}{Initial \protect\ac{LIGO}}
\newacronym{eLIGO}{eLIGO}{Enhanced \protect\ac{LIGO}}
\newacronym{LHO}{LHO}{\protect\ac{LIGO} Hanford Observatory}
\newacronym{LLO}{LLO}{\protect\ac{LIGO} Livingston Observatory}
\newacronym{LSC}{LSC}{\protect\ac{LIGO} Scientific Collaboration}
\newacronym{CBC}{CBC}{compact binary coalescence}
\newacronym{GW}{GW}{gravitational wave}
\newacronym{DQ}{DQ}{data quality}
\newacronym{SNR}{SNR}{signal-to-noise ratio}
\newacronym{S4}{S4}{Science Run 4}
\newacronym{S5}{S5}{Science Run 5}
\newacronym{S6}{S6}{Science Run 6}
\newacronym{HEPI}{HEPI}{hydraulic external pre-isolator}
\newacronym{KW}{KW}{Kleine-Welle}
\newacronym{hveto}{HVeto}{HierarchichalVeto}
\newacronym{PEM}{PEM}{physical environment monitor}
\newacronym{DetChar}{DetChar}{the Detector Characterisation working group of the \protect\ac{LSC}}
\newacronym{LVEA}{LVEA}{Large Vacuum Equipment Area}
\newacronym{OMC}{OMC}{output mode cleaner}
\newacronym{NPRO}{NPRO}{non-planar ring oscillator}
\newacronym{USDOE}{USDOE}{United States Department of Energy}
\newacronym{PN}{PN}{post-Newtonian}
\newacronym{ISCO}{ISCO}{innermost stable circular orbit}
\newacronym{PSD}{PSD}{power spectral density}
\newacronym{TT}{TT}{transverse-traceless}
\newacronym{DARM}{DARM}{differential arm}
\newacronym{CARM}{CARM}{common arm}
\newacronym[longplural={input test masses}]{ITM}{ITM}{input test mass}
\newacronym[longplural={end test masses}]{ETM}{ETM}{end test mass}
\newacronym{BS}{BS}{beam splitter}
\newacronym{PRC}{PRC}{power recycling cavity}
\newacronym{PRCL}{PRCL}{the \protect\acl{PRC} length\protect\glsunset{PRC}}
\newacronym{PRM}{PRM}{power recycling mirror}
\newacronym{SRC}{SRC}{signal recycling cavity}
\newacronym{SRM}{SRM}{signal recycling mirror}
\newacronym{MICH}{MICH}{the short Michelson interferometer formed by the \protect\ac{BS} and \acp{ITM}}
\newacronym{GR}{GR}{general relativity}
\newacronym{SR}{SR}{special relativity}
\newacronym{GPS}{GPS}{Global Positioning System}
\newacronym{VSR}{VSR}{Virgo Science Run}
\newacronym{GRB}{GRB}{gamma ray burst}
\newacronym{GEODC}{GEODC}{GEO600 detector characterisation}
\newacronym{GBM}{GBM}{Gamma-ray Burst Monitor}
\newacronym{GWB}{GWB}{gravitational-wave burst}
\newacronym{GEO-HF}{GEO-HF}{GEO High Frequency}
\newacronym{PyLAL}{PyLAL}{Python \protect\ac{LIGO} Algorithm Library}
\newacronym{AEI}{AEI}{Albert Einstein Institute}
\newacronym{FFT}{FFT}{fast Fourier transform}
\newacronym{EM}{EM}{electromagnetic}
\newacronym{HEN}{HEN}{high-energy neutrino}
\newacronym{BAT}{BAT}{Burst Alert Telescope}
\newacronym{EFE}{EFE}{Einstein Field Equation}
\newacronym{MC}{MC}{mode cleaner}
\newacronym{CMB}{CMB}{cosmic microwave background}
\newacronym{USA}{USA}{United States of America}
\newacronym{ASD}{ASD}{amplitude spectral density}
\newacronym{CW}{CW}{continuous wave}
\newacronym{SGWB}{SGWB}{stochastic \protect\ac{GW} background}
\newacronym{TCS}{TCS}{thermal compensation system}
\newacronym[longplural={figures of merit}]{FOM}{FOM}{name={figure of merit}}
\newacronym{VME}{VME}{Versa Module Europa}
\newacronym{REFL}{REFL}{reflected}
\newacronym{RBS}{RBS}{REFL beam servo}
\newacronym{EDR}{EDR}{efficiency-to-deadtime ratio}
\newacronym{BNS}{BNS}{binary neutron star}
\newacronym{DMT}{DMT}{Data Monitoring Tool}
\newacronym{cwb}{cWB}{Coherent WaveBurst}
\newacronym{etg}{ETG}{event trigger generator}
\newacronym{UPV}{UPV}{Used Percentage Veto}
\newacronym{ODC}{ODC}{Online Detector Characterisation}
\newacronym{TEM}{TEM}{transverse \protect\acl{EM}}
\newacronym{QPD}{QPD}{quadrant photo-diode}
\newacronym{ADC}{ADC}{analog-to-digital converter}
\newacronym{RF}{RF}{radio-frequency}
\newacronym{SFT}{SFT}{short Fourier transform}
\newacronym{AC}{AC}{alternating current}
\newacronym{DAC}{DAC}{digital-to-analog converter}
\newacronym{BCV}{BCV}{bilinear-coupling veto}
\newglossaryentry{DC readout}{name={DC readout}, description={homodyne detection}}
\newglossaryentry{glitch}{name={glitch}, description={transient noise event}, plural={glitches}, descriptionplural={transient noise events}}
\newglossaryentry{test mass}{name={test mass}, description={core interferometer optic}, descriptionplural={core interferometer optics}, plural={test masses}}
\newglossaryentry{veto}{name={veto}, plural={vetoes}, description={distinct method of identifying times of questionable \protect\ac{DQ}}, descriptionplural={distinct methods of identifying times of questionable \protect\ac{DQ}}}
\newglossaryentry{veto segment}{name={veto segment}, description={segment of science-quality data removed by a \protect\gls{veto}}, descriptionplural={segments of science-quality data removed by a \protect\gls{veto}}}
\newglossaryentry{flag}{name={\protect\ac{DQ} flag}, description={unique marker associated with a \protect\glspl{veto}}}
\newglossaryentry{hwinj}{name={hardware injection}, description={simulated \protect\ac{GW} signal injected into the instrument by exciting one of the \protect\glspl{test mass}}, descriptionplural={simulated \protect\ac{GW} signals injected into the instrument by exciting one of the \protect\glspl{test mass}}}
\newglossaryentry{line}{name={spectral line}, description={long-duration, narrow-bandwidth noise peak}, descriptionplural={long-duration, narrow-bandwidth noise peaks}}
\newglossaryentry{violin mode}{name={violin mode}, description={resonance of the \protect\gls{test mass} suspensions}, descriptionplural={resonances of the \protect\glspl{test mass} suspensions}}
\newglossaryentry{power line}{name={power line}, description={harmonic of the 60\,Hz mains power supply}, descriptionplural={harmonics of the 60\,Hz mains power supply}}
\newglossaryentry{deadtime}{name={deadtime}, description={the fractional amount of analysis time that has been vetoed}}
\newglossaryentry{efficiency}{name={efficiency}, description={the fractional number of \protect\ac{GW} candidate events removed by a \protect\gls{veto}}}
\newglossaryentry{trigger}{name={trigger}, description={event produced by a search algorithm}, descriptionplural={events produced by a search algorithm}}

\glsdisablehyper

\usepackage{xspace}
\newcommand\OmegaPipeline{$\Omega$-pipeline\xspace}

\begin{document}
    \title[LIGO Detector Characterization in S6]%
          {Characterization of the LIGO detectors during their sixth %
           science run}
    \author{%
J~Aasi$^{1}$,
J~Abadie$^{1}$,
B~P~Abbott$^{1}$,
R~Abbott$^{1}$,
T~Abbott$^{2}$,
M~R~Abernathy$^{1}$,
T~Accadia$^{3}$,
F~Acernese$^{4,5}$,
C~Adams$^{6}$,
T~Adams$^{7}$
% ...
}

\address {$^{1}$LIGO - California Institute of Technology, Pasadena, CA 91125, USA }
\address {$^{2}$Louisiana State University, Baton Rouge, LA 70803, USA }
\address {$^{3}$Laboratoire d'Annecy-le-Vieux de Physique des Particules (LAPP), Universit\'e de Savoie, CNRS/IN2P3, F-74941 Annecy-le-Vieux, France }
\address {$^{4}$INFN, Sezione di Napoli, Complesso Universitario di Monte S.Angelo, I-80126 Napoli, Italy }
\address {$^{5}$Universit\`a di Salerno, Fisciano, I-84084 Salerno, Italy }
\address {$^{6}$LIGO - Livingston Observatory, Livingston, LA 70754, USA }
\address {$^{7}$Cardiff University, Cardiff, CF24 3AA, United Kingdom }
% ...

    \begin{abstract}
In 2009-2010, the \ac{LIGO} operated together with international partners Virgo and GEO600 as a network to search for gravitational waves of astrophysical origin.
The sensitivity of these detectors was limited by a combination of noise sources inherent to the instrumental design and its environment, often localized in time or frequency, that couple into the gravitational-wave readout.
Here we review the performance of the LIGO instruments during this epoch, the work done to characterize the detectors and their data, and the effect that transient and continuous noise artefacts have on the sensitivity of LIGO to a variety of astrophysical sources.
\end{abstract}

    \glsresetall

    \section{Introduction}
\label{sec:introduction}
Between July 2009 and October 2010, the \ac{LIGO}~\cite{Abbott:2007kv} operated two 4-kilometre laser interferometers as part of a global network aiming to detect and study \acp{GW} of astrophysical origin.
These detectors, at \acl{LHO}, WA (\acs{LHO}\glsunset{LHO}), and \acl{LLO}, LA (\acs{LLO}\glsunset{LLO}) -- dubbed `H1' and `L1', and operating beyond their initial design with greater sensitivity -- took data during \ac{S6} in collaboration with GEO600~\cite{Grote:2008zz} and Virgo~\cite{Acernese:2008zza}.

The data from each of these detectors have been searched for \ac{GW} signals from a number of sources, including \acp{CBC}~\cite{Abadie:2011np,Abadie:2012gt,Aasi:2013km}, generic short-duration \ac{GW} bursts~\cite{Abadie:2012gt,Abadie:2012rq}, non-axisymmetric spinning neutron stars~\cite{Aasi:2012fw}, and a \ac{SGWB}~\cite{Abbott:2009ws}.
The performance of each of these analyses is measured by the searched volume of the universe multiplied by the searched time duration; however, long and short duration artefacts in real data, such as narrow-bondwidth noise lines and glitches, further restrict the sensitivity of GW searches.

Searches for transient \ac{GW} signals including \acp{CBC} and \ac{GW} bursts are sensitive to many short-duration noise events (glitches\glsunset{glitch}), coming from a number of environmental, mechanical, and electronic mechanisms that are not fully understood.
Each search pipeline employs signal-based methods to distinguish a \ac{GW} event from noise based on knowledge of the expected waveform~\cite{Klimenko:2008fu,Sutton:2009gi,Harry:2010fr,Babak:2012zx}, but also relies on careful studies of the detector behaviour to provide information that leads to improved data quality through `vetoes' that remove data likely to contain noise artefacts.
Searches for long-duration \acp{CW} and a \ac{SGWB} are sensitive to disturbances from spectral lines and other sustained noise artefacts.
These effects cause elevated noise at a given frequency and so impair any search over these data.

This paper describes the work done to characterize the \ac{LIGO} detectors and their data during \ac{S6}, and estimates the increase in sensitivity for analyses resulting from detector improvements and data quality vetoes.
This work follows from previous studies of \ac{LIGO} data quality during \ac{S5}~\cite{Blackburn:2008ah,Slutsky:2010ff} and \ac{S6}~\cite{Christensen:2010zz,McIver:2012ky}.
Similar studies have also been performed for the Virgo detector relating to data taking during \acp{VSR} 2, 3 and 4~\cite{Robinet:2010zz,Aasi:2012wd}.

\Cref{sec:detectors} details the configuration of the LIGO detectors during S6, and \cref{sec:overview} details their performance over this period, outlining some of the problems observed and improvements seen.
\Cref{sec:issues} describes examples of important noise sources that were identified at each site and steps taken to mitigate them.
In \cref{sec:vetoes}, we present the performance of data-quality vetoes when applied to each of two astrophysical data searches: the ihope \ac{CBC} pipeline~\cite{Babak:2012zx} and the \ac{cwb} burst pipeline~\cite{Klimenko:2008fu}.
A short conclusion is given in \cref{sec:conclusion}, along with plans for characterization of the next-generation \acl{aLIGO} detectors, currently under construction.

    \section{Configuration of the LIGO detectors during the sixth science run}
\label{sec:detectors}
\begin{figure}
    \centering
    \includegraphics[width=\textwidth]{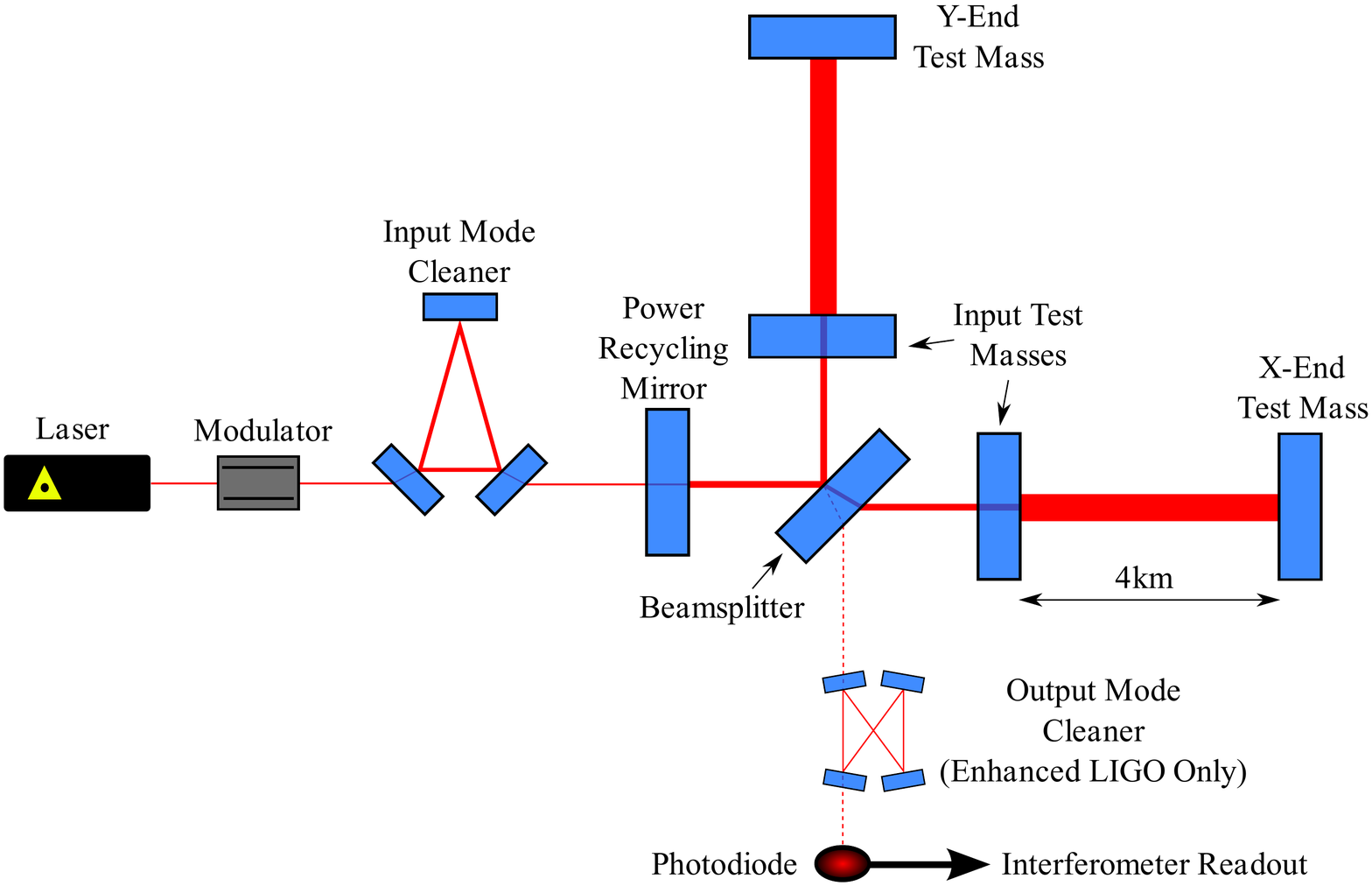}
    \caption[Optical layout of the LIGO interferometers during S6.]%
            {Optical layout of the \protect\ac{LIGO} interferometers during %
             \protect\ac{S6}~\cite{Smith:2009bx}. The layout differs from %
             that used in \protect\ac{S5} with the addition of the output %
             mode cleaner.}
    \label{fig:layout}
\end{figure}

The first-generation \ac{LIGO} instruments were versions of a Michelson interferometer~\cite{Michelson:1887wc} with Fabry-Perot arm cavities, with which \ac{GW} amplitude is measured as a strain of the 4-kilometre arm length, as shown in \cref{fig:layout}~\cite{Smith:2009bx}.
In this layout, a diode-pumped, power-amplified Nd:YAG laser generated a carrier beam in a single longitudinal mode at $1064\,$nm~\cite{Savage:1998uv}.
This beam passed through an electro-optic modulator which added a pair of \ac{RF} sidebands used for sensing and control of the test mass positions, before the modulated beam entered a triangular optical cavity.
This cavity (the `input mode cleaner') was configured to filter out residual higher-order spatial modes from the main beam before it entered the main interferometer.

The conceptual Michelson design was enhanced with the addition of input test masses at the beginning of each arm to form Fabry-Perot optical cavities.
These cavities increase the storage time of light in the arms, effectively increasing the arm length.
Additionally, a power-recycling mirror was added to reflect back light returned towards the input, equivalent to increasing the input laser power.
During \ac{S5}, the relative lengths of each arm were controlled to ensure that the light exiting each arm cavity interfered destructively at the output photodiode, and all power was returned towards the input.
In such `dark fringe' operation, the phase modulation sidebands induced in the arms by interaction with \acp{GW} would interfere constructively at the output, recording a \ac{GW} strain in the demodulated signal.
In this configuration, the \ac{LIGO} instruments achieved their design sensitivity goal over the 2-year \ac{S5} run.
A thorough description of the initial design is given in \cite{Abbott:2007kv}.

For \ac{S6} a number of new systems were implemented to improve sensitivity and to prototype upgrades for the second-generation \ac{aLIGO} detectors~\cite{Smith:2009bx,Harry:2010zz}.
The initial input laser system was upgraded from a 10\,W output to a maximum of 35\,W, with the installation of new master ring oscillator and power amplifier systems~\cite{Adhikari:2006tt}.
The higher input laser power from this system improved the sensitivity of the detectors at high frequencies ($>150\,$Hz) and allowed prototyping of several key components for the \ac{aLIGO} laser system~\cite{Kwee:2012db}.
Additionally, an improved CO$_2$-laser thermal-compensation system was installed~\cite{Ballmer:2005up,Amin:2010in} to counteract thermal lensing caused by expansion of the test mass coating substrate due to heat from absorption of the main beam.

An alternative \ac{GW} detection system was installed, replacing the initial heterodyne readout scheme~\cite{Sigg:2001vn}.
A special form of homodyne detection, known as \textit{DC readout}, was implemented, whereby the interferometer is operated slightly away from the dark fringe.
In this system, \ac{GW}-induced phase modulations would interfere with the main beam to produce power variations on the output photodiode, without the need for demodulating the output signal.
In order to improve the quality of the light incident on the output photodiode in this new readout system, an \ac{OMC} cavity was installed to filter out the higher-order mode content of the output beam~\cite{SmithLefebvre:2011fp}, including the \ac{RF} sidebands.
The \ac{OMC} was required to be in-vacuum, but also highly stable, and so a single-stage prototype of the new \ac{aLIGO} two-stage seismic isolation system was installed for the output optical platform~\cite{Bertolini:2006bl}, from which the \ac{OMC} was suspended.

Futhermore, controls for seismic feed-forward to a hydraulic actuation system were improved at \ac{LLO} to combat the higher level of seismic noise at that site~\cite{DeRosa:2012bf}.
This system used signals from seismometers at the Michelson vertex, and at ends of each of the arms, to suppress the effect of low-frequency ($\mathord{\lesssim}$ 10 Hz) seismic motion on the instrument.

    \section{Detector sensitivity during S6}
\label{sec:overview}
The maximum sensitivity of any \ac{GW} search, such as those cited in \cref{sec:introduction}, is determined by the amount of coincident multi-detector operation time and astrophysical reach of each detector.
In searches for transient signals these factors determine the number of sources that could be detected during a science run, while in those for continuous signals they determine the accumulated signal power over that run.

The \ac{S6} run took place between July 7th 2009 and October 20th 2010, with each detector recording over seven months of data in that period.
The data-taking was split into four epochs, A--D, identifying distinct analysis periods set by changes in detector performance or the detector network itself.
Epochs A and B ran alongside the second \acl{VSR}\glsunset{VSR} (\acs{VSR}2) before that detector was taken off-line for a major upgrade~\cite{Aasi:2012wd}.
\ac{S6}A ran for $\mathord{\sim}2$ months before a month-long instrumental commissioning break, and \ac{S6}B ran to the end of 2009 before another commissioning break.
The final 2 epochs, C and D, spanned a continuous period of detector operation, over nine months in all, with the distinction marking the start of \ac{VSR}3 and the return of a three-detector network.

\begin{figure}
    \centering
    \includegraphics[width=1.0\textwidth]{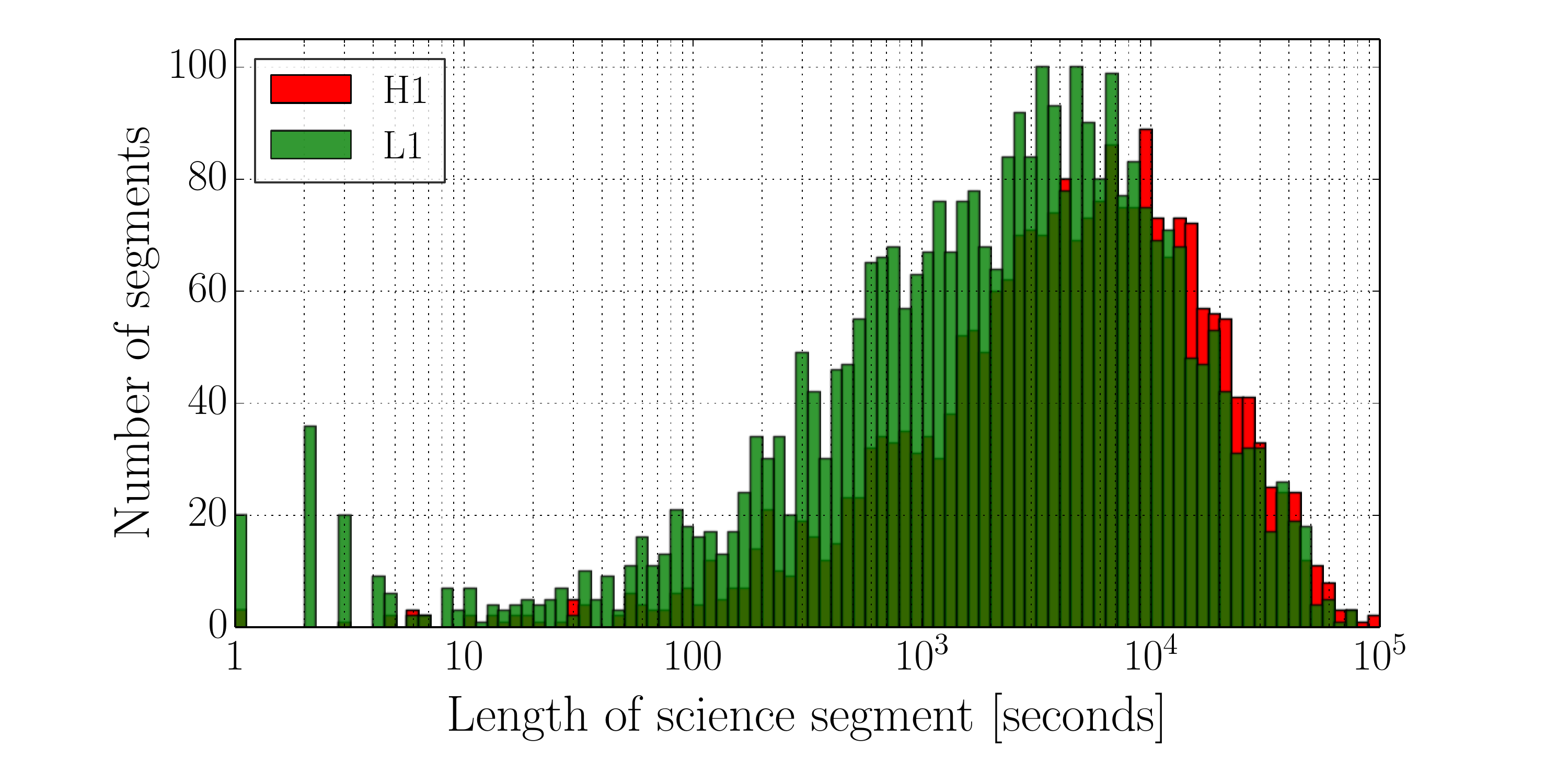}
    \caption[Histogram of science segment duration for the LIGO detectors %
             during S6]%
            {A histogram of the duration of each science segment for the %
             \protect\ac{LIGO} detectors during \protect\ac{S6}. %
             The distribution is centred around $\mathord{\sim} 1$ hour.}
    \label{fig:seghist}
\end{figure}

Instrumental stability over these epochs was measured by the detector duty factor -- the fraction of the total run time during which science-quality data was recorded.
Each continuous period of operation is known as a \textit{science segment}, defined as time when the interferometer is operating in a nominal state and the spectral sensitivity is deemed acceptable by the operator and scientists on duty.
A science segment is typically ended by a critically large noise level in the instrument at which time interferometer control cannot be maintained by the electronic control system (known as \textit{lock-loss}).
However, a small number of segments are ended manually during clean data in order to perform scheduled maintenance, such as a calibration measurement.
\Cref{fig:seghist} shows a histogram of science segment duration over the run.
The majority of segments span several hours, but there are a significant number of shorter segments, symptomatic of interferometer instability.
In particular, for L1 the number of shorter segments is higher than that for H1, a result of poor detector stability during the early part of the run, especially during \ac{S6}B.

\Cref{tab:segsummary} summarises the science segments for each site over the four run epochs.
\begin{table}
    \centering
    \subfloat[H1 (\protect\acl{LHO})\label{subtab:h1_segs}]{
        \begin{tabular}{p{0.8cm}||p{2.1cm}|p{2.2cm}|p{1.9cm}|p{2cm}}
            \scriptsize Epoch & \scriptsize Median duration (mins) &%
            \scriptsize Longest duration (hours) &%
            \scriptsize Total live time (days) & \scriptsize Duty factor (\%)\\
            \hline
            \hline
            S6A & 54.0 & 13.4 & 27.5 & 49.1\\
            \hline
            S6B & 75.2 & 19.0 & 59.2 & 54.3\\
            \hline
            S6C & 82.0 & 17.0 & 82.8 & 51.4\\
            \hline
            S6D & 123.4 & 35.2 & 74.7 & 63.9\\
            \hline
        \end{tabular}
    }\\
    \subfloat[L1 (\protect\acl{LLO})\label{subtab:l1_segs}]{
        \begin{tabular}{p{0.8cm}||p{2.1cm}|p{2.2cm}|p{1.9cm}|p{2cm}}
            \hline
            S6A & 39.3 & 11.8 & 25.6 & 45.7\\
            \hline
            S6B & 17.3 & 21.3 & 40.0 & 38.0\\
            \hline
            S6C & 67.5 & 21.4 & 82.3 & 51.1\\
            \hline
            S6D & 58.2 & 32.6 & 75.2 & 64.3\\
            \hline
        \end{tabular}
    }
    \caption{Science segment statistics for the \protect\ac{LIGO} detectors %
             over the four epochs of \protect\ac{S6}.}
    \label{tab:segsummary}
\end{table}
Both sites saw an increase in duty factor, that of H1 increasing by $\mathord{\sim}15$ percentage points, and L1 by nearly 20 between epochs A and D.
Additionally, the median duration of a single science-quality data segment more than doubled at both sites between the opening epochs (S6A and S6B) and the end of the run.
These increases in stability highlight the developments in understanding of the critical noise couplings~\cite{Abbott:2007kv} and how they affect operation of the instruments (see \cref{sec:issues} for some examples), as well as improvements in the control system used to maintain cavity resonance.

\begin{figure}
    \centering
    \includegraphics[width=1.0\textwidth]{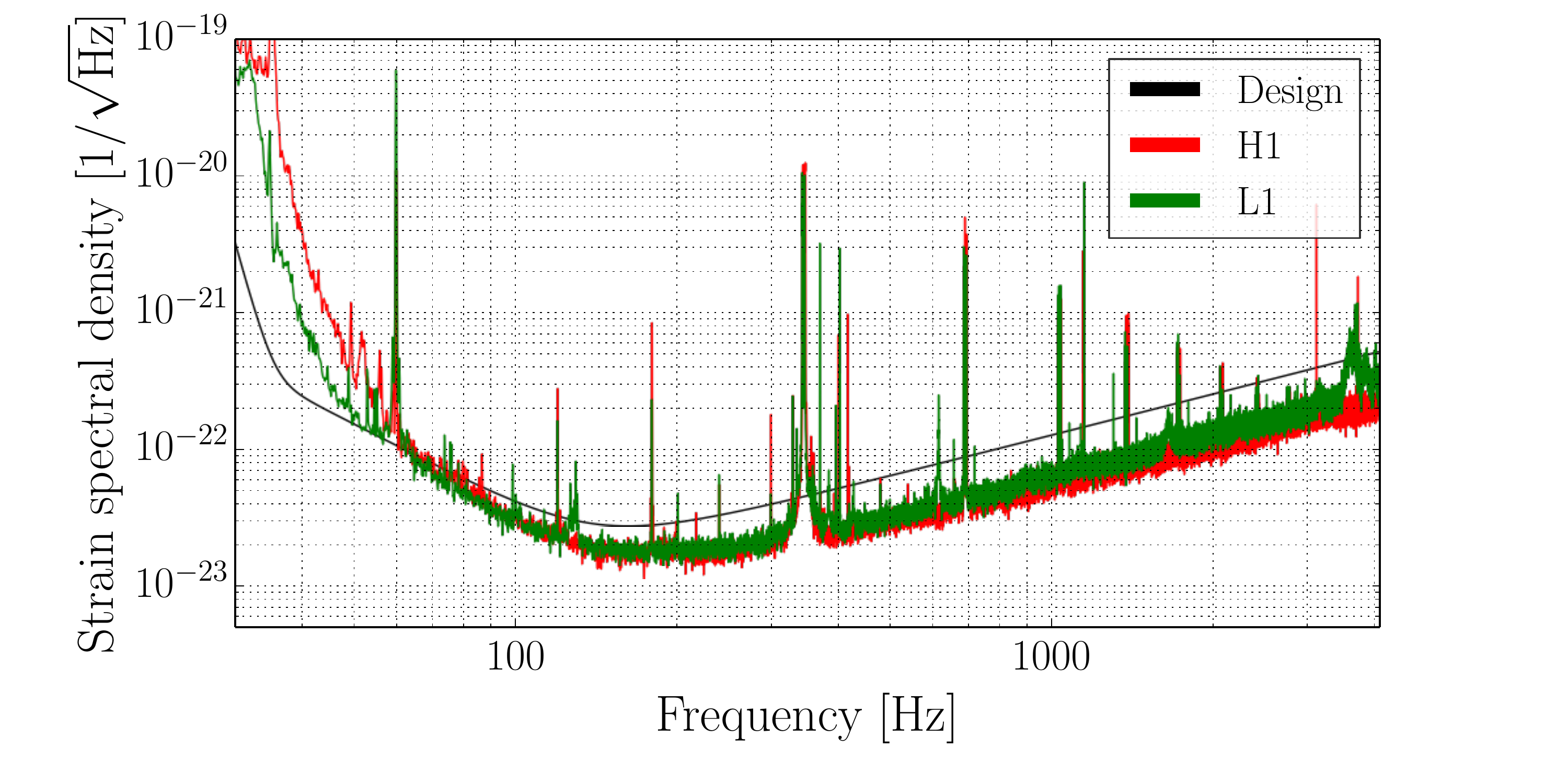}
    \caption[Typical strain amplitude sensitivity of the LIGO %
             detectors during S6]%
            {Typical strain amplitude sensitivity of the LIGO %
             detectors during S6.}
    \label{fig:asd}
\end{figure}

The sensitivity to \acp{GW} of a single detector is typically measured as a strain amplitude spectral density of the calibrated detector output.
This is determined by a combination of noise components, some fundamental to the design of the instruments, and some from additional noise coupling from instrumental and environmental sources.
\Cref{fig:asd} shows the typical amplitude spectral densities of the \ac{LIGO} detectors during \ac{S6}.
The dominant contribution below 40\,Hz is noise from seismically-driven motion of the key interferometer optics, and from the servos used to control their alignment.
The reduced level of the seismic wall at L1 relative to H1 can be, in part, attributed to the prototype hydraulic isolation installed at that observatory~\cite{DeRosa:2012bf}.
Intermediate frequencies, 50-150\,Hz, have significant contributions from Brownian motion -- mechanical excitations of the test masses and their suspensions due to thermal energy~\cite{Levin:1998es,Harry:2006gd} -- however, some of the observed limiting noise in this band was never understood.
Above 150\,Hz, shot noise due to variation in incident photon flux at the output port is the dominant fundamental noise source~\cite{Buonanno:2001bp}. 
The sensitivity is also limited at many frequencies by narrow-band line structures, described in detail in \cref{subsec:lines}.
The spectral sensitivity gives a time-averaged view of detector performance, and so is sensitive to the long-duration noise sources and signals, but rather insensitive to transient events.

\begin{figure}
    \centering
    \includegraphics[width=1.0\textwidth]{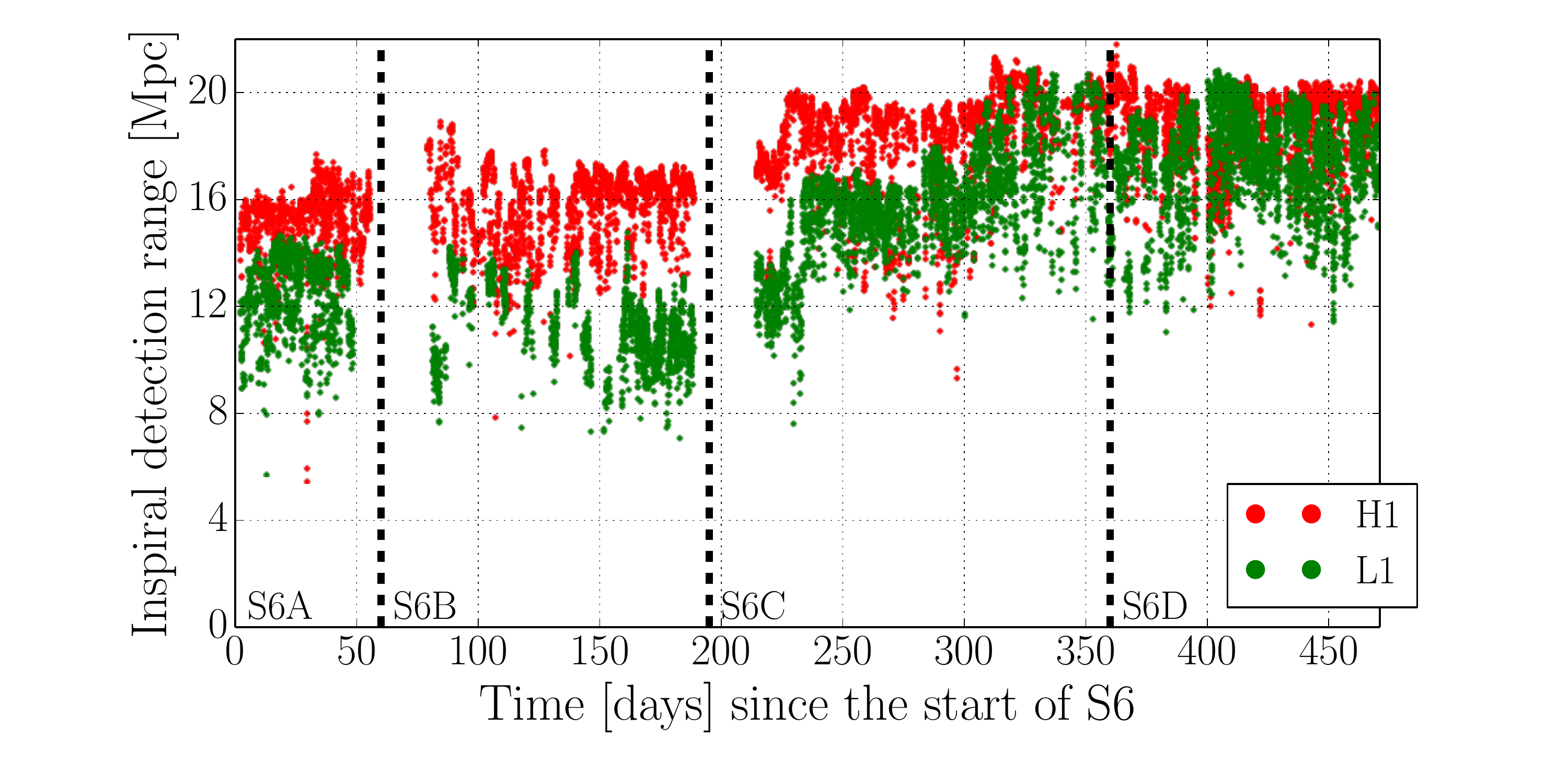}
    \caption[The inspiral detection range of the LIGO detectors during S6]%
            {The inspiral detection range of the \protect\ac{LIGO} detectors %
             throughout \protect\ac{S6} to a binary neutron star merger, %
             averaged over sky location and orientation. %
             The rapid improvements between epochs %
             can be attributed to hardware and control changes implemented %
             during commissioning periods.}
    \label{fig:range}
\end{figure}

A standard measure of a detector's astrophysical reach is the distance to which that instrument could detect \ac{GW} emission from the inspiral of a \acl{BNS} system with a \ac{SNR} of 8~\cite{Finn:1992xs,LIGO:2012aa}, averaged over source sky locations and orientations.
\Cref{fig:range} shows the evolution of this metric over the science run, with each data point representing an average over 2048\,s of data.
Over the course of the run, the detection range of H1 increased from $\mathord{\sim}16$ to $\mathord{\sim}20$\,Mpc, and of L1 from $\mathord{\sim}14$ to $\mathord{\sim} 20$\,Mpc.
The instability of \ac{S6}B at L1 can be seen between days 80--190, with a lower duty factor (also seen in \cref{tab:segsummary}) and low detection range; this period included extensive commissioning of the seismic feed-forward system at \ac{LLO}~\cite{DeRosa:2012bf}.

The combination of increased amplitude sensitivity and improved duty factor over the course of \ac{S6} meant that the searchable volume of the universe for an astrophysical analysis was greatly increased.

    \section{Data-quality problems in S6}
\label{sec:issues}
While the previous section described the performance of the \ac{LIGO} detectors over the full span of the S6 science run, there were a number of isolated problems that had detrimental effects on the performance of each of the observatories at some time.
Each of these problems, some of which are detailed below, introduced excess noise at specific times or frequencies that hindered astrophysical searches over the data.

Under ideal conditions, all excess noise sources can be quickly identified in the experimental set-up and corrected, either with a hardware change, or a modification of the control system.
However, not all such fixes can be implemented immediately, or at all, and so noisy periods in auxiliary data (other data streams not directly associated with gravitational-wave readout) must be noted and recorded as likely to adversely affect the \ac{GW} data.
During \ac{S6}, these \ac{DQ} flags and their associated time segments were used by analysis groups to inform decisions on which data to analyse, or which detection candidates to reject as likely noise artefacts, the impact of which will be discussed in \cref{sec:vetoes}.

The remainder of this section details a representative set of specific issues that were present for some time during \ac{S6} at \ac{LHO} or \ac{LLO}, some of which were fixed at the source, some which were identified but could not be fixed, and one which was never identified.


\subsection{Seismic noise}
\label{subsec:seismic}
Throughout the first-generation \ac{LIGO} experiment, the impact of seismic noise was a fundamental limit to the sensitivity to \acp{GW} below 40\,Hz.
However, throughout \ac{S6} (and earlier science runs), seismic noise was also observed to be strongly correlated with transient noise glitches in the detector output, not only at low frequencies, but also at much higher frequencies ($\mathord{\sim}100\mbox{-}200\,$Hz).

\begin{figure}
    \centering
    \includegraphics[width=\textwidth]{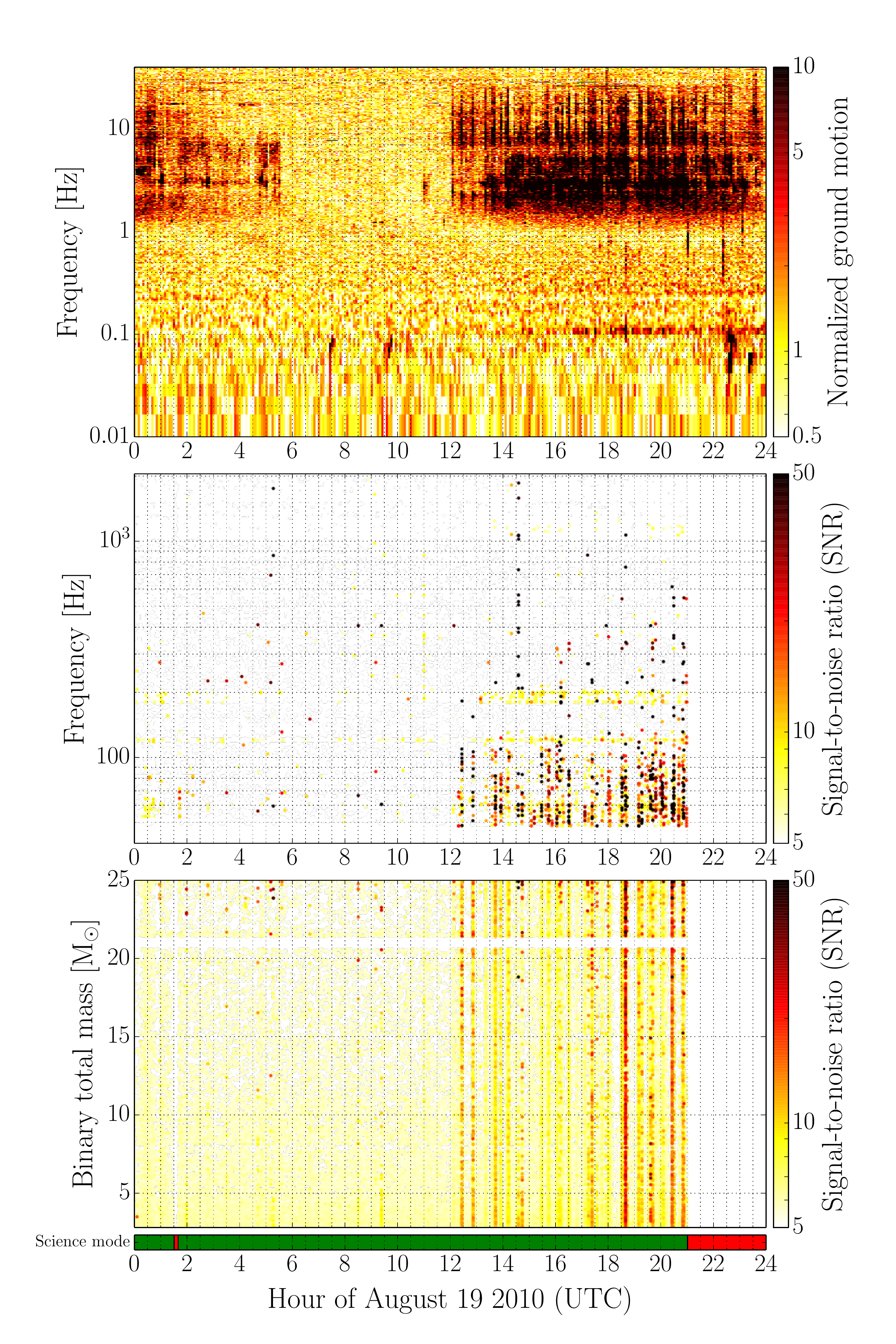}
    \caption[Seismic noise at LHO]%
            {Seismic motion of the laboratory floor at LHO %
             (normalized, top) and its correlation into %
             \protect\ac{GW} burst (middle) and inspiral (bottom) analyses.}
    \label{fig:seisveto}
\end{figure}

The top panels of \cref{fig:seisveto} show the seismic ground motion at \ac{LHO}, both in specific frequency bands (left) and as seen by the \OmegaPipeline transient search algorithm~\cite{Chatterji:2005t} (right) - this panel shows localised seismic noise events in time and frequency coloured by their \ac{SNR}.
The lower left and right panels show transient events in the gravitational-wave strain data as recorded by single-interferomter burst and \ac{CBC} inspiral searches respectively.
Critically, during periods of high seismic noise, the inspiral analysis `daily ihope'~\cite{Babak:2012zx} produced candidate event triggers across the full range of signal templates, severely limiting the sensitivity of that search.

While great efforts were made to reduce the coupling of seismic noise into the interferometer~\cite{DeRosa:2012bf}, additional efforts were required to improve the identification of loud transient seismic events that were likely to couple into the \ac{GW} readout~\cite{MacLeod:2011up}.
Such times were recorded and used by astrophysical search groups to veto candidate events from analyses, proving highly effective in reducing the noise background of such searches.

\subsection{Seismically-driven length-sensing glitches}
While transient seismic noise was a problem throughout the science run, during late 2009 the presence of such noise proved critically disruptive at \ac{LLO}.
During \ac{S6}B, the majority of \glspl{glitch} in L1 were correlated with noise in the length control signals of two short length degrees of freedom: \ac{PRCL}, and the short Michelson formed by the beam-splitter and the input test masses (\acs{MICH}\glsunset{MICH}).
Both of these length controls were glitching simultaneously, and these glitches were correlated with more than 70\% of the glitches in the \ac{GW} data.

It was discovered that high microseismic noise was driving large instabilities in the \acl{PRC} that caused significant drops in the circulating power, resulting in large glitches in both the \ac{MICH} and \ac{PRC} length controls.
These actuation signals, applied to the main interferometer optics, then coupled into the detector output.

%
This issue was eliminated via commissioning of a seismic feed-forward system~\cite{DeRosa:2012bf} that decreased the \ac{PRC} optic motion by a factor of three.
The glitchy data before the fix were identified by both the \ac{hveto} and \ac{UPV} algorithms~\cite{Smith:2011an,Isogai:2010} -- used to rank auxiliary signals according to the statistical significance of \gls{glitch} coincidence with the \ac{GW} data -- with those times used by the searches to dismiss noise artefacts from their results (more in \cref{sec:vetoes}).

\subsection{Upconversion of low-frequency noise due to the Barkhausen effect}
\label{subsec:upconversion}
\begin{figure}
    \centering
    \includegraphics[width=\textwidth]{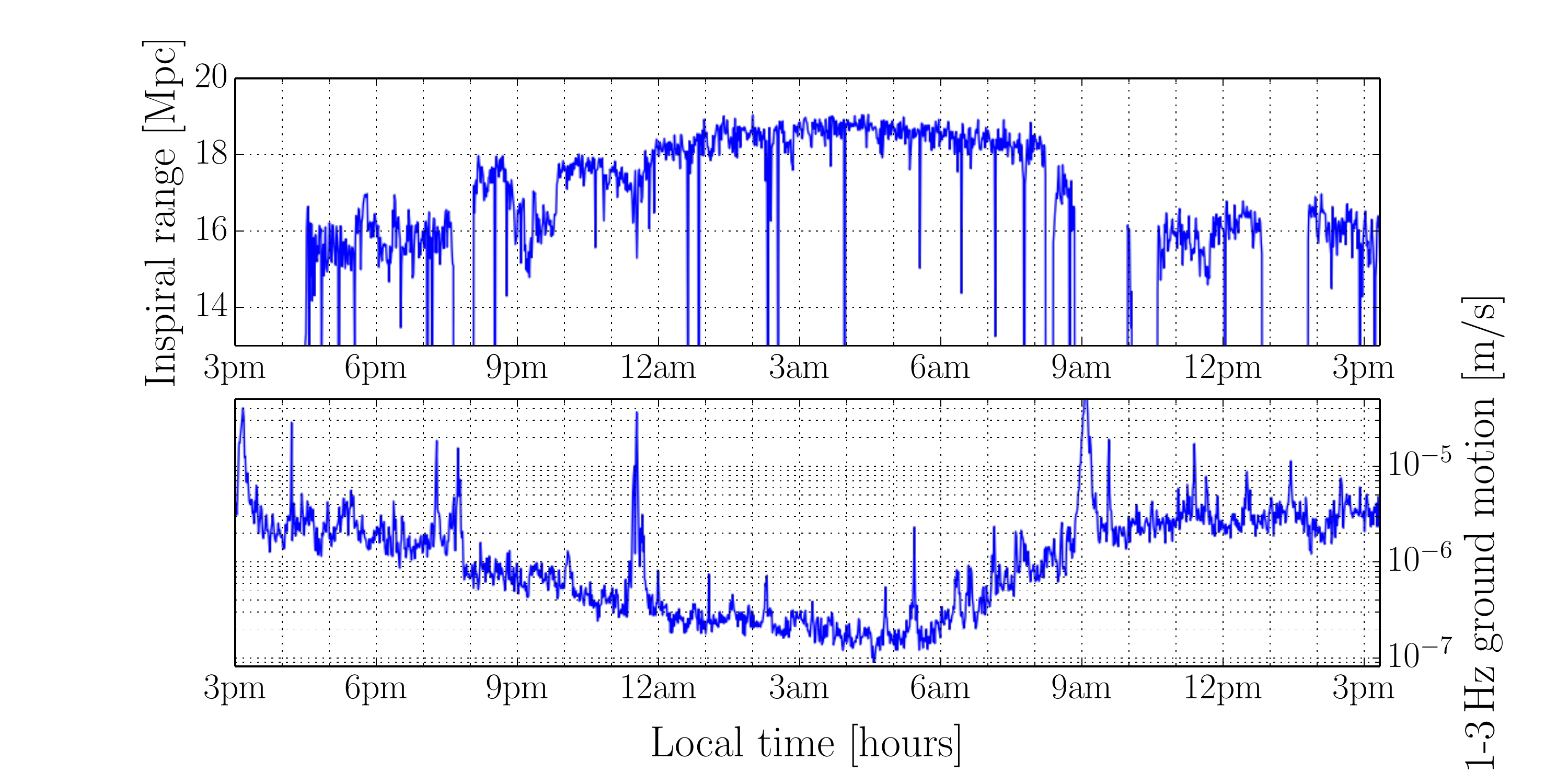}
    \caption[Sensitive distance to a binary neutron star (top) and ground %
             motion in the 1--3\,Hz band (bottom) for a day at \protect\ac{LLO}]
            {Sensitive distance to a binary neutron star (top) and ground %
             motion in the 1--3\,Hz band (bottom) for a day at %
             \protect\ac{LLO}. The inverse relationship is believed to be due %
             to non-linear upconversion of low frequency seismic ground %
             motions to higher frequency ($\mathord{\sim}40-200\,$Hz) noise %
             in the GW output.}
    \label{fig:upconversion}
\end{figure}

In earlier science runs, as well as affecting performance below 40\,Hz, increased levels of ground motion below 10\,Hz had been associated with increases in noise in the $40$--$200$\,Hz band.
This noise, termed seismic upconversion noise, was produced by passing trucks, distant construction activities, seasonal increases in water flow over dams, high wind, and earthquakes~\cite{Daw:2004qd,Smith:2009bx,Slutsky:2010ff,MacLeod:2011up}.
During \ac{S6}, this noise was often the limiting noise source at these higher frequencies.
\Cref{fig:upconversion} shows a reduction in the sensitive range to \ac{BNS} inspirals, contemporaneous with the workday increase in anthropogenic seismic noise.

Experiments subsequently showed that seismic upconversion noise levels correlated better with the amplitudes of the currents to the electromagnets that held the test masses in place as the ground moved than with the actual motion of the test masses or of the ground.
An empirical, frequency-dependent function was developed to estimate upconversion noise from the low-frequency test mass actuation currents.
This function was used to produce flags that indicated time periods that were expected to have high levels of seismic upconversion noise.

\begin{figure}
    \centering
    \includegraphics[width=.8\textwidth]{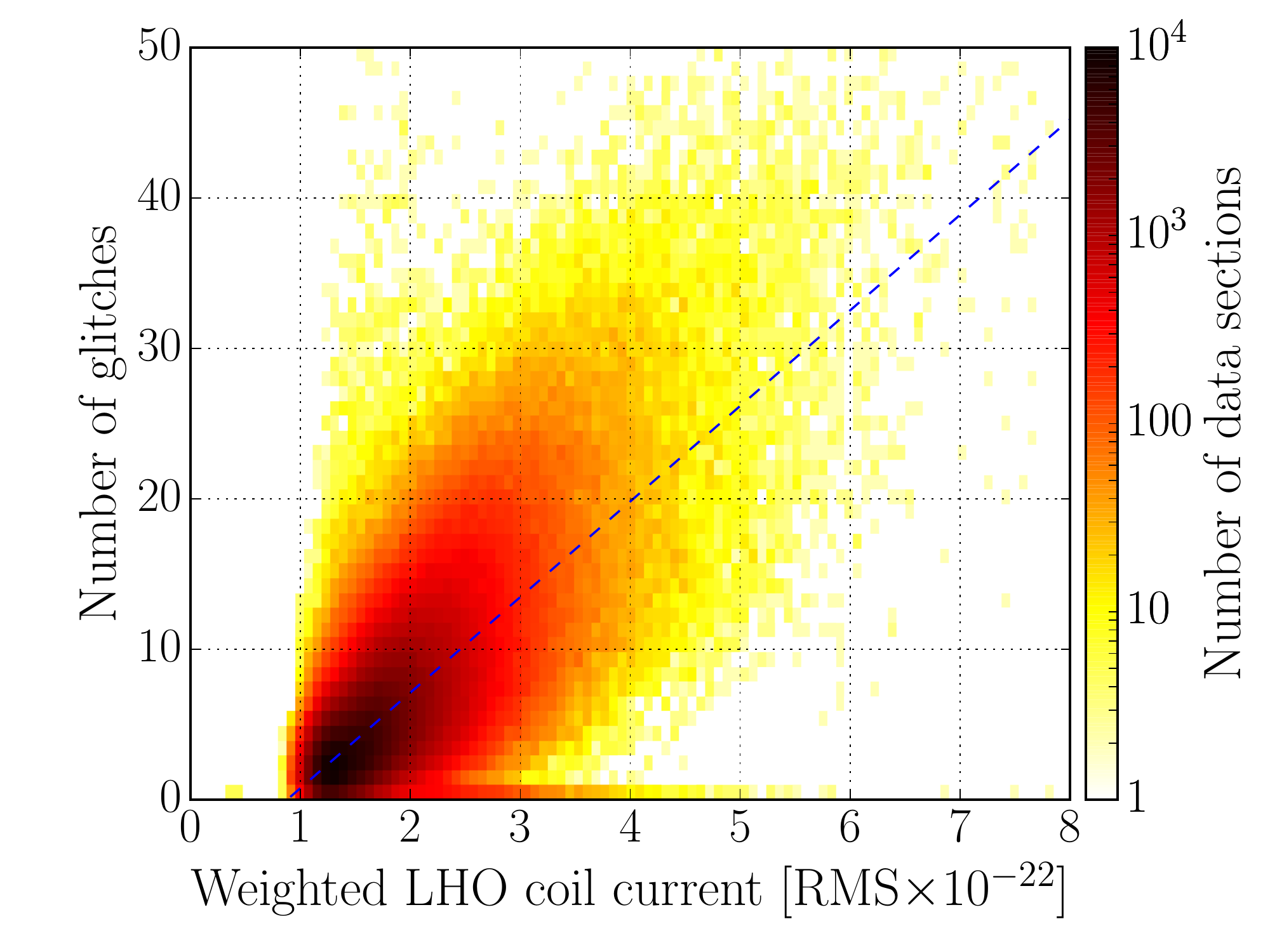}
    \caption[Correlation between low \protect\ac{SNR} \glspl{glitch} in the %
             \protect\ac{GW} data, and current in the test mass coil %
             at H1]
            {Correlation between low \protect\ac{SNR} \glspl{glitch} in the %
             \protect\ac{GW} data, and current in the test mass coil %
             at H1. %
             The $r^2$ value measures the normalized cross-correlation %
             through a least-squares fit.
             This correlation is indicative of the Barkhausen effect.}
    \label{fig:coil_current}
\end{figure}

In addition to average reductions in sensitivity, upconverted seismic noise transients further reduced sensitivity to unmodelled \ac{GW} bursts.
\Cref{fig:coil_current} shows that the rate of low-\ac{SNR} glitches in the \ac{GW} data -- in a frequency band above that expected from linear seismic noise coupling -- was correlated with the test mass actuation current, suggesting that seismic upconversion was the source of a low-\ac{SNR} noise background that limited \ac{GW} burst detection.

Investigations found that seismic upconversion noise bursts were clustered in periods of high slope in the amplitude of the magnetic actuator current.
This was evidence that the seismic upconversion noise was Barkhausen noise~\cite{Manson:2002co}: magnetic field fluctuations produced by avalanches of magnetic domains in ferromagnetic materials that occur when the domains align with changing magnetic fields.
The Barkhausen noise hypothesis was supported by investigations in which the noise spectrum was reproduced by magnetic fields that were generated independently of the system.

These investigations also suggested that the putative source of the Barkhausen noise was near or inside the test mass actuators.
It was originally thought that the source of this upconversion noise was Barkhausen noise from NbFeB magnets, but a swap to less noisy SmCo magnets did not significantly reduce the noise~\cite{Weiss:2008cc}.
However, it was found that fasteners inside the magnetic actuator, made of grade 303 steel, were ferromagnetic, probably because they were shaped or cut when cold.
For \ac{aLIGO}, grade 316 steel, which is much less ferromagnetic after cold working, is being used at the most sensitive locations.

\subsection{Beam jitter noise}
\label{subsec:jitter}
As described previously, one of the upgrades installed prior to \ac{S6} was the \acl{OMC}, a bow-tie-shaped cavity designed to filter out higher-order modes of the main laser beam before detection at the output photodiode.
As known from previous experiments at GEO600~\cite{Prijatelj:2012hu}, the mode transmission of this cavity is very sensitive to angular fluctuations of the incident beam, whereby misalignment of the beam would cause non-linear power fluctuations of the transmitted light~\cite{SmithLefebvre:vo,Fricke:2011dv}.

At \ac{LIGO}, low-frequency seismic noise and vibrations of optical tables were observed to mix with higher-frequency beam jitter on the \ac{OMC} to produce noise sidebands around the main jitter frequency.
The amplitude of these sidebands was unstable, changing with the amount of alignment offset, resulting in transient noise at these frequencies, the most sensitive region of the \ac{LIGO} spectrum, as seen in \cref{fig:seisveto} (bottom left panel).
Mitigation of these glitches involved modifications of the suspension system for the auxiliary optics steering the beam into the \ac{OMC}, to minimise the coupling of optical table motion to beam motion.
Additionally, several other methods were used to mitigate and control beam jitter noise throughout the run: full details are given in \cite{Fricke:2011dv}.

\subsection{Mechanical glitching at the reflected port}
While the problems described up to this point have been inherent to the design or construction of either interferometer, the following two issues were both caused by electronics failures associated with the \ac{LHO} interferometer.

The first of these was produced by faults in the servo actuators used to stabilize the pointing of the beam at the reflected port of the interferometer.
This position is used to sense light reflected from the \ac{PRC} towards the input, and generate control signals to correct for arm-cavity motion.
The resulting \glspl{glitch} coupled strongly into the gravitational-wave data at $\mathord{\sim} 37$\,Hz and harmonics.

\begin{figure}
    \includegraphics[width=\linewidth]{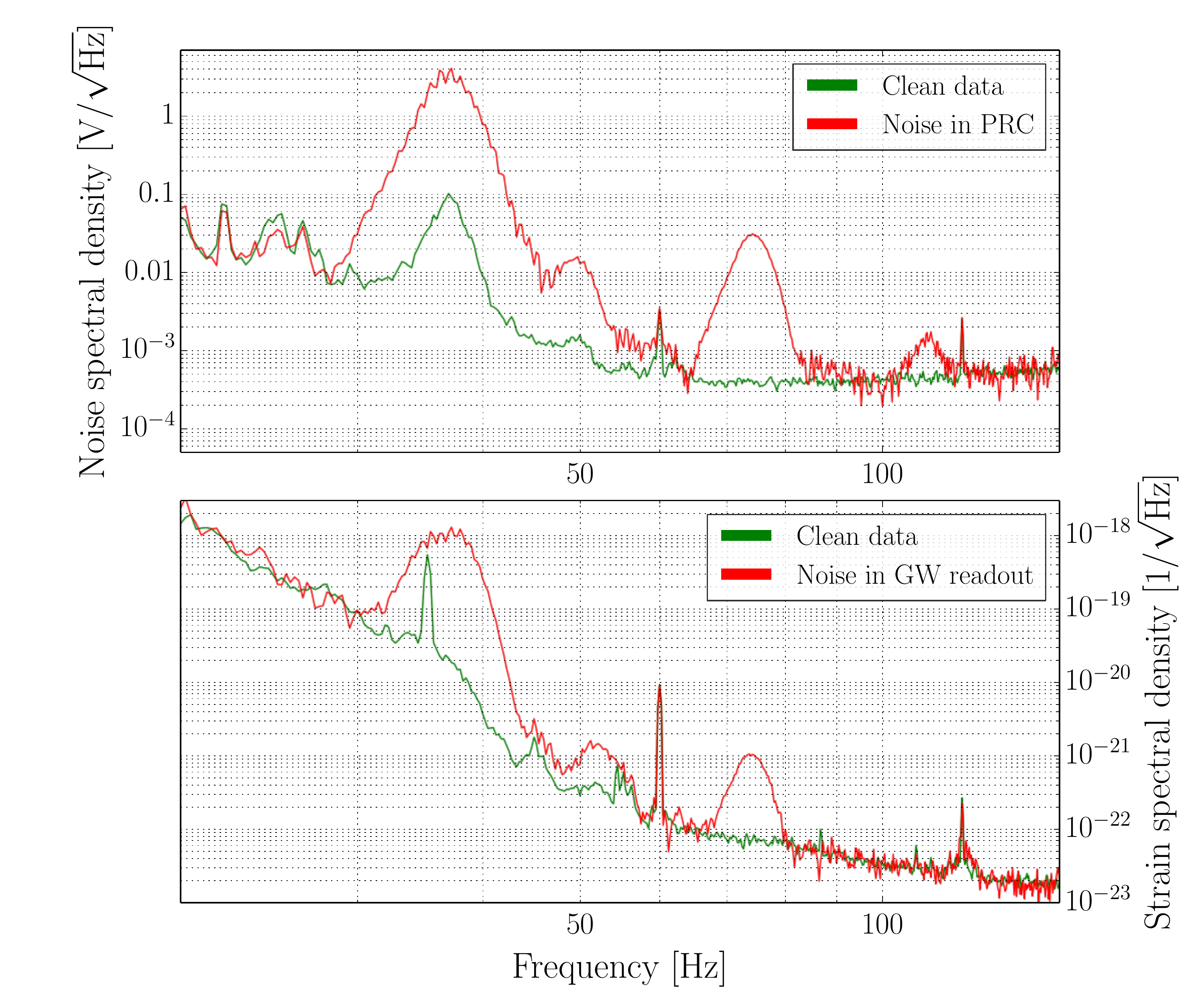}
    \caption[Noise peaks caused by a faulty power supply]%
            {Broad noise peaks centred at 37\,Hz and its harmonics in the %
             \protect\acl{PRC} length signal (top) and the GW %
             output error signal (bottom).
             Each panel shows the spectrum as a noisy period (red) in %
             comparison with a reference taken from clean data (green).}
    \label{fig:rbs}
\end{figure}

The source of the glitches was identified with the help of \ac{hveto}, which discovered that a number of angular and length sensing channels derived from photodiodes at the reflected port were strongly coupled with events in the \ac{GW} data.
Figure \ref{fig:rbs} shows the broad peaks in the spectra of one length sensing channel and the un-calibrated \ac{GW} readout compared to a quiet reference time.
On top of this, accelerometer signals from the optical table at the reflected port were found to be coupling strongly, having weak but coincident glitches.

These accelerometer coincidences indicated that the glitches were likely produced by mechanical motions of steering mirrors resulting from a faulty piezoelectric actuation system.
Because of this, this servo was decomissioned for the rest of the run, leading to an overall improvement in data quality.

\subsection{Broadband noise bursts from poor electrical connections}
\label{subsec:grid_glitches}
\begin{figure}
    \centering
    \includegraphics[width=\textwidth]{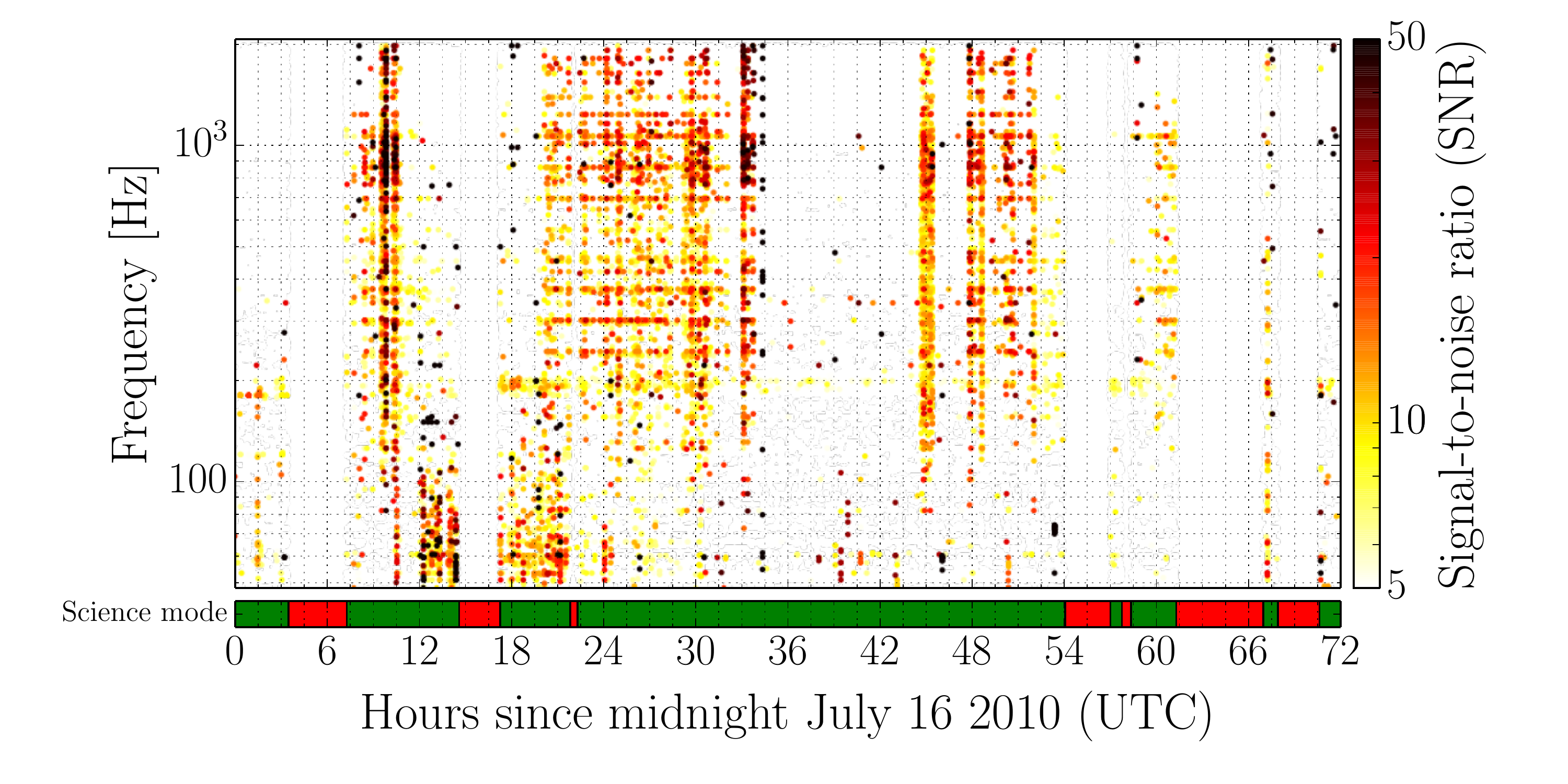}
    \caption[Noise events identified by the \OmegaPipeline due to %
             faulty electronics]%
            {Noise events in the \protect\ac{GW} strain data recorded by %
             the \OmegaPipeline over a 60-hour period at \protect\ac{LHO}. %
             The high \protect\ac{SNR} events above 100\,Hz in hours 7--10, %
             20-34, and 44-42, were caused by %
             broadband noise from a faulty electrical connection. %
             The grid-like nature of these events is due to the discrete %
             tiling in frequency by the trigger generator.}
    \label{fig:grid}
\end{figure}

The second of the electronics problems caused repeated, broadband glitching in the \ac{LHO} \ac{GW} readout towards the end of \ac{S6}.
Periods of glitching would last from minutes to hours, and greatly reduced the instrumental sensitivity over a large frequency range, as shown in \cref{fig:grid}.

The main diagnostic clues were coincident, but louder, glitches in a set of \acp{QPD} sensing beam motion in the \ac{OMC}.
It was unlikely that these sensors could detect a glitch in the beam more sensitively than the \ac{GW} readout photo-diode, and so the prime suspect then became the electronics involved with recording data from these \acp{QPD}.

In the process of isolating the cause, several other electronics boards in the output mode cleaner were inspected, re-soldered, and swapped for spares.
The problem was finally solved by re-soldering the connections on the electronics board that provided the high-voltage power supply to drive a piezoelectric transducer.

\subsection{Spectral lines}
\label{subsec:lines}
Just as searches for transient signals are limited by instrumental glitches, so too our searches for steady signals are limited by a number of instrumental narrow-band peaks representing specific frequencies at which noise was elevated for a significant amount of time, in many cases for the entire science run.
Many spectral lines are fundamental to the design and operation of the observatories, including \ac{AC} power lines from the U.S. mains supply, at 60\,Hz; violin modes from core-optic suspensions, around 350\,Hz; and various calibration lines used to measure the interferometer response function.

Each of these features can be seen in \cref{fig:asd} at their fundamental frequency and a number of harmonics; however, also seen are a large number of lines from unintended sources, such as magnetic and vibrational couplings.
These noise lines can have a damaging effect on any search for \acp{GW} if the frequencies of the incoming signal and of the lines overlap for any time; this is especially troublesome for searches for continuous gravitational-wave emitters.

\begin{figure}
    \includegraphics[width=\linewidth]{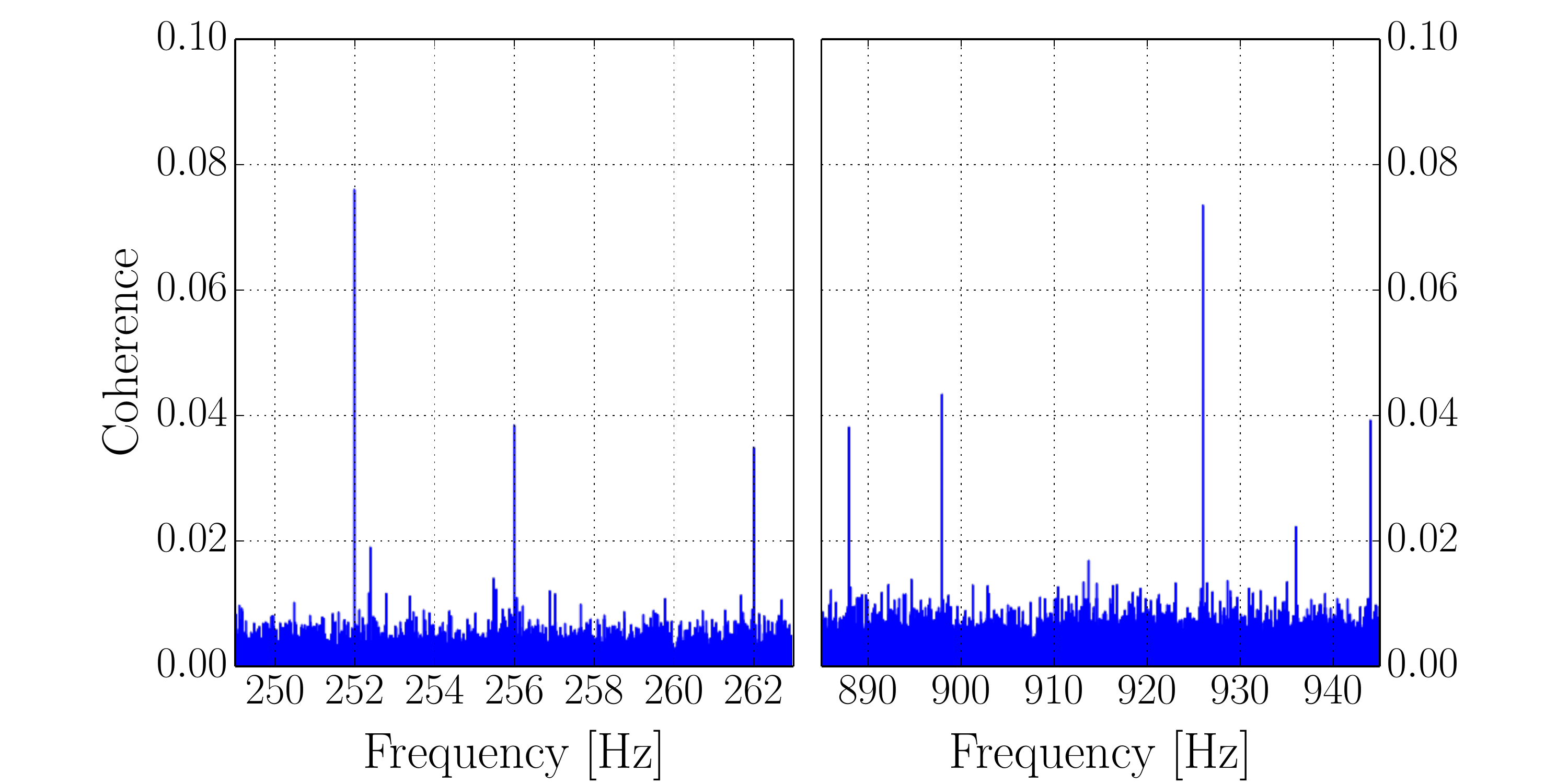}
    \caption[The coherence between the L1 GW readout signal and %
             data from a magnetometer in the central building at LLO.]
            {The coherence between the L1 \protect\ac{GW} readout signal %
             and data from a magnetometer in the central building at %
             \protect\ac{LLO} over one week of March 2010.
             2\,Hz and 16\,Hz harmonics were seen to be coherent at numerous %
             locations across the operating band of both interferometers, %
             affecting the sensitivity of long-duration \protect\ac{GW} %
             searches.}
    \label{fig:mag_coherence}
\end{figure}

Throughout \ac{S6}, series of lines were seen at both observatories as 2\,Hz and 16\,Hz harmonics.
\Cref{fig:mag_coherence} shows two separate groups of peaks in these harmonic sets found in coherence between the \ac{GW} data for L1 and a magnetometer located near the output photo-detector.
These lines were a serious concern for both the \ac{CW} and \ac{SGWB} searches due to their appearance at both observatories~\cite{Coughlin:2011pb}, leading to contamination of the coincidence-based searches for \ac{CW} sources.
Commissioning work on the \ac{OMC} control system at \ac{LHO} in early 2010 suggested that the use of a specific set of high-frequency ($>1$\,kHz) dither signals resulted in low-frequency noise generated by the \acp{DAC}, but although resulting modifications to the controls scheme removed the 2\,Hz series at that site, this hypothesis was never confirmed.

A number of other lines were isolated at either observatory site~\cite{Coughlin:2011pb}, and while not discussed in detail here, the cumulative effect of all spectral lines on searches for long-duration gravitational-wave sources is discussed in detail in \cref{sec:vetoes}.

\subsection{The `spike' glitch}
\begin{figure}
    \centering
    \includegraphics[width=\linewidth]{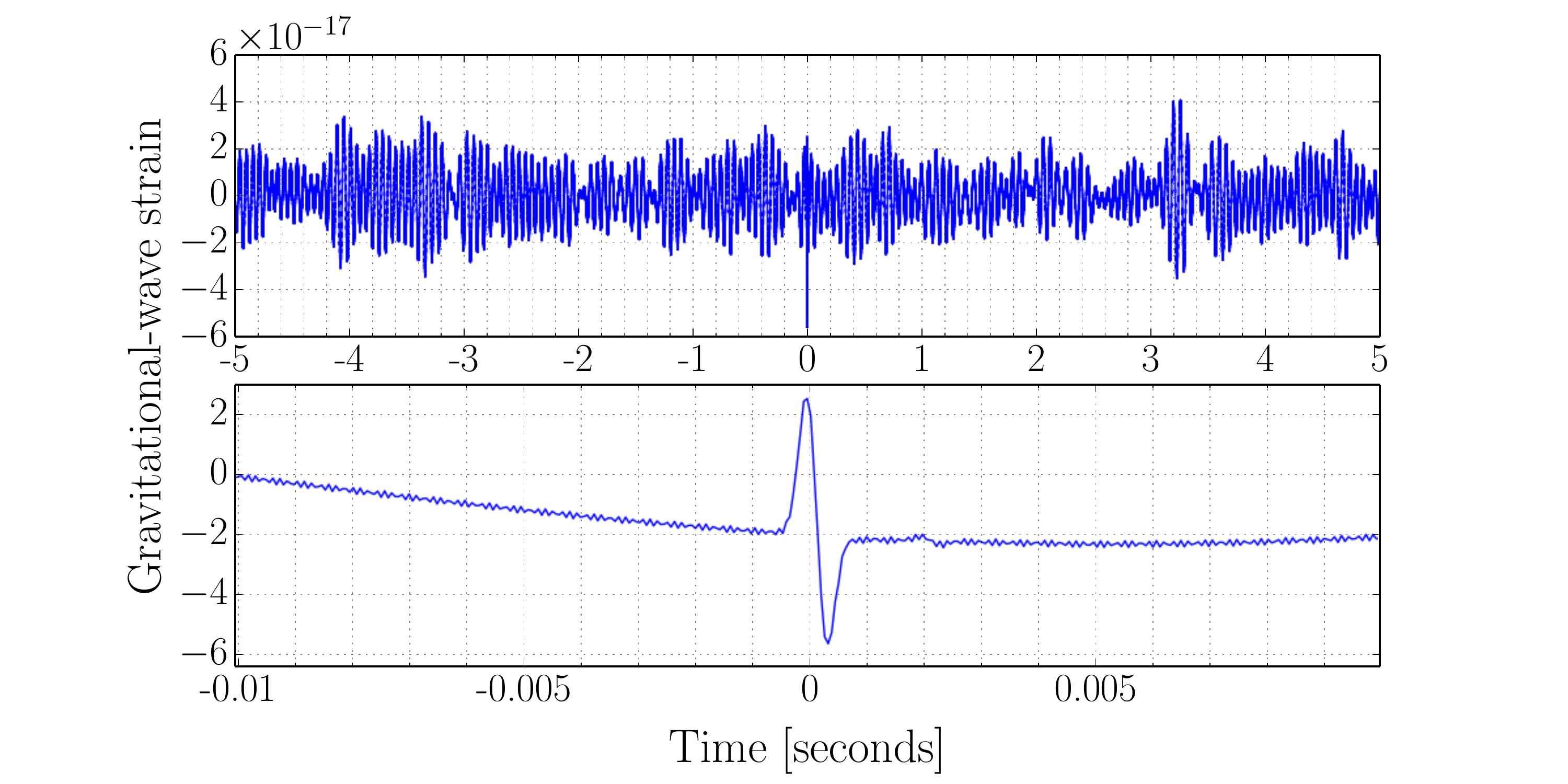}
    \caption[A spike glitch in the raw \ac{GW} photodiode signal for L1]%
            {A spike glitch in the raw \ac{GW} photodiode signal for L1. %
             The top panel shows the glitch in context with 10\,s of data, %
             while the bottom shows the glitch profile as described in the %
             text.}
    \label{fig:spike}
\end{figure}

The spike glitch was the name given to a class of very loud transients seen in the L1 instrument.
They were characterized by a distinctive shape in the time series of the signal on the \ac{GW} output photodiode, beginning with a rapid but smooth dip (lasting $\mathord{\sim} 1\,$ms) before a period of damped oscillation lasting $\mathord{\sim} 3$ milliseconds, as shown in Figure \ref{fig:spike}.
The amplitude of these glitches was extremely large, often visible in the raw time-series (which is normally dominated by low-frequency seismic motion), with the \OmegaPipeline typically resolving these events with \acp{SNR} ranging from 200 to well over 20,000.

The size and rapidity of the initial glitch suggested that the source was after the beams had re-combined at the beam-splitter before detection at the readout photodiode.
The damped oscillations after the initial dip, however, were likely due to the response of the length control loop of the interferometer, meaning an actual or apparent sudden dip in the light on the output photodiode could explain the entire shape of the spike glitch.
To investigate this possibility, the interferometer was run in a configuration where the light did not enter the arm cavities, but went almost directly into the \ac{OMC}, removing the length and angular control servos from consideration.
Sharp downward dips in the light were seen during this test, although they were 0.2 milliseconds wide, much narrower than the initial dips of the spike glitches.

Despite this investigation and many others, the cause of the spike glitch was never determined.
However, these glitches were clearly not of astrophysical origin, and were not coherent with similar events in H1, allowing the \ac{CBC} signal search to excise them from analyses by vetoing time around glitches detected in L1 with unreasonably high \ac{SNR}.
For future science runs, \acl{aLIGO} will consist of almost entirely new hardware, so whether the spike glitch or something very similar will be seen in new data remains to be seen.

    \section{The impact of data quality on gravitational wave searches}
\label{sec:vetoes}
The impact of non-Gaussian, non-stationary noise in the LIGO detectors on searches for \acp{GW} is significant.
Such loud glitches, such as the spike glitch, can mask or greatly disrupt transient \ac{GW} signals present in the data at the same time, while high rates of lower \ac{SNR} glitches can significantly increase the background in searches for these sources.
Additionally, spectral lines and continued glitching in a given frequency range reduces the sensitivity of searches for long-duration signals at those frequencies.
Both long- and short-duration noise sources have a notable effect on search sensitivity if not mitigated.

Non-Gaussian noise in the detector outputs that can be correlated with auxiliary signals that have negligible sensitivity to \acp{GW} can be used to create flags for noisy data; these flags can then be used in astrophysical searches to remove artefacts and improve sensitivity.
With transient noise, the flags are used to identify time segments in which the data may contain \glspl{glitch}.
Similarly, spectral lines are recorded as frequencies, or narrow frequency bands, at which the detector sensitivity is reduced.


\subsection{Data quality vetoes for transient searches}
\label{subsec:vetoes_transient}
In this section, the impact of noisy data is measured by its effect on the primary analyses of the LIGO-Virgo transient search groups~\cite{Abadie:2011np,Abadie:2012rq}:
\begin{itemize}
\item the low-mass \ac{CBC} search `ihope'~\cite{Babak:2012zx} is a coincidence-based analysis in which data from each detector are filtered against a bank of binary inspiral template signals, producing an \ac{SNR} time-series for each.
Peaks in \ac{SNR} across multiple detectors are considered coincident if the separation in time and matched template masses are small~\cite{Robinson:2008un}.
This analysis also uses a $\chi^2$-statistic test to down-rank signals with high \ac{SNR} but a spectral shape significantly different to that of the matched template~\cite{Allen:2004gu}.
\item the all-sky \ac{cwb} algorithm~\cite{Klimenko:2008fu} calculates a multi-detector statistic by clustering time-frequency pixels with significant energy that are coherent across the detector network. 
\end{itemize}
In both cases, the multi-detector events identified are then subject to a number of consistency tests before being considered detection candidates.

The background of each search is determined by relatively shifting the data from multiple detectors in time.
These time shifts are much greater than the time taken for a \ac{GW} to travel between sites, ensuring that any multi-detector events in these data cannot have been produced by a single astrophysical signal.

Although both searches require signal power in at least two detectors, strong \glspl{glitch} in a single detector coupled with Gaussian noise in others still contributed significantly to the search background during \ac{S6}.
\ac{DQ} flags were highly effective in removing these noise artefacts from the analyses.
The effect of a time-domain \ac{DQ} flag can be described by its \textit{deadtime}, the fraction of analysis time that has been vetoed; and its \textit{efficiency}, the fractional number of \ac{GW} candidate events removed by a veto in the corresponding deadtime.

Flag performances are determined by their \ac{EDR}; random flagging and vetoing of data gives \acs{EDR} $\mathord{\simeq} 1$, whereas effective removal of \acp{glitch} gives a much higher value.
Additionally, the \emph{used percentage} -- the fraction of auxiliary channel glitches which coincide with a \ac{GW} candidate event -- allows a measure of the strength of the correlation between the auxiliary and GW channel data.

Each search group chose to apply a unique set of \ac{DQ} flags in order to minimise deadtime whilst maximising search sensitivity; for example, the \ac{CBC} search teams did not use a number of flags correlated with very short, high-frequency disturbances, as these do not trigger their search algorithm, while these flags were used in searches for unmodelled \ac{GW} bursts.

We present the effect of three categories of veto on each of the above searches in terms of reduction in analysable time and removal of noise artefacts from the search backgrounds.
Only brief category definitions are given, for full descriptions see \cite{Slutsky:2010ff}.

\subsubsection{Category 1 vetoes.}
\label{subsubsec:category_1}
The most egregious interferometer performance problems are flagged as category 1.
These flags denote times during data taking when the instrument was not running under the designed configuration, and so should not be included in any analysis.

The \ac{DMT} automatically identified certain problems in real time, including losses of cavity resonance, and errors in the $h(t)$ calibration.
Additionally, scientists monitoring detector operation in the control room at each observatory manually flagged individual time segments that contained observed instrumental issues and errors.

All \ac{LIGO}-Virgo search groups used category 1 vetoes to omit unusable segments of data; as a result their primary effect was in the reduction in analysable time over which searches were performed.
This impact is magnified by search requirements on the duration for analysed segments, with the \ac{cwb} and ihope searches requiring a minimum of 316 and 2064 seconds of contiguous data respectively.
\begin{table}
    \centering
    \begin{tabular}{c||l|l||l|l||}
       & \multicolumn{2}{c||}{Absolute deadtime \% (seconds)} & %
       \multicolumn{2}{c||}{Search deadtime \% (seconds)}\\
       Instrument & \protect\ac{cwb} & ihope & \protect\ac{cwb} & ihope \\
       \hline\hline
       H1 & 0.3\% (53318) & 0.4\% (176079) & 0.4\% (77617) &  3.8\% (786284)\\
       L1 & 0.4\% (75016) & 0.1\% (20915) & 0.7\% (137115) & 6.2\% (1180976)\\
    \end{tabular}
    \caption{Summary of the reduction in all time and analysable time by %
             category 1 veto segments during \protect\ac{S6}}
    \label{tab:cat1_deadtime}
\end{table}
\Cref{tab:cat1_deadtime} outlines the absolute deadtime (fraction of science-quality data removed) and the search deadtime (fractional reduction in analysable time after category 1 vetoes and segment selection).
At both sites the amount of science-quality time flagged as category 1 is less than half of one percent, highlighting the stability of the instrument and its calibration. However, the deadtime introduced by segment selection is significantly higher, especially for the \ac{CBC} analysis.
The long segment duration requirement imposed by the ihope pipeline results in an order of magnitude increase in search deadtime relative to absolute deadtime.

\subsubsection{Categories 2 and 3.}
The higher category flags were used to identify likely noise artefacts.
Category 2 veto segments were generated from auxiliary data whose correlation with the \ac{GW} readout has been firmly demonstrated by instrumental commissioning and investigations.
Category 3 includes veto segments from less well understood statistical correlations between noisy data in an auxiliary channel and the \ac{GW} readout.
Both the ihope and cWB search pipelines produce a first set of candidate event triggers after application of category 2 vetoes, and a reduced set after application of category 3.

The majority of category 2 veto segments were generated in low-latency by the \ac{DMT} and include things like photodiode saturations, digital overflows, and high seismic and other environmental noise.
At category 3, the \ac{hveto}~\cite{Smith:2011an}, \ac{UPV}~\cite{Isogai:2010}, and \ac{BCV}~\cite{Ajith:2014aea} algorithms were used, by the burst and \ac{CBC} analyses respectively, to identify coupling between auxiliary data and the \ac{GW} readout.

\Cref{tab:cat23_deadtime} gives the absolute, relative, and cumulative deadtimes of these categories after applying category 1 vetoes and segment selection criteria, outlining the amount of analysed time during which event triggers were removed.
As with category 1, category 2 vetoes have deadtime $\mathcal{O}(1)\%$, but with significantly higher application at L1 compared to H1.
This is largely due to one flag vetoing the final 30 seconds before a lock-loss combined with the relative abundance of short data-taking segments for L1.
Additionally, photodiode saturations and computational timing errors were more prevalent at the \ac{LLO} site than at \ac{LHO} and so contribute to higher relative deadtime.

Category 3 flags contributed $\mathcal{O}(10)\%$ deadtime for each instrument.
While this level of deadtime is relatively high, as we shall see, the efficiency of these flags in removing background noise events makes such cuts acceptable to the search groups.

\begin{table}[th]
    \centering
    \begin{tabular}{rc||l|l||l|l||}
       & & \multicolumn{2}{c||}{H1} & \multicolumn{2}{c||}{L1}\\
       Deadtime type & Cat. & \protect\ac{cwb} & ihope & %
                               \protect\ac{cwb} & ihope \\
       \hline\hline
       \multirow{2}{*}{Absolute \% (s)} &%
           2 & 0.26\% & 0.77\% & 1.59\% & 1.53\%\\
       \cline{2-6}
         & 3 & 7.90\% & 9.26\% & 8.54\% & 7.03\%\\
       \hline
       Relative \% (s) & 3 & 7.73\% & 9.00\% & 7.06\% & 6.10\%\\
       \hline
       Cumulative \% (s) & 3 & 7.97\% & 9.71\% & 8.54\% & 7.54\%\\
    \end{tabular}
    \caption[Summary of the absolute, relative, and cumulative deadtimes %
             introduced by category 2 and 3 veto segments during %
             \protect\ac{S6}]%
            {Summary of the absolute, relative, and cumulative deadtimes %
             introduced by category 2 and 3 veto segments during %
             \protect\ac{S6}. %
             The relative deadtime is the additional time removed by %
             category 3 not vetoed by category 2, and cumulative deadtime
             gives the total time removed from the analysis.}
    \label{tab:cat23_deadtime}
\end{table}

\begin{figure}[t]
    \centering
    \subfloat[H1]{%
        \includegraphics[width=.35\textwidth]{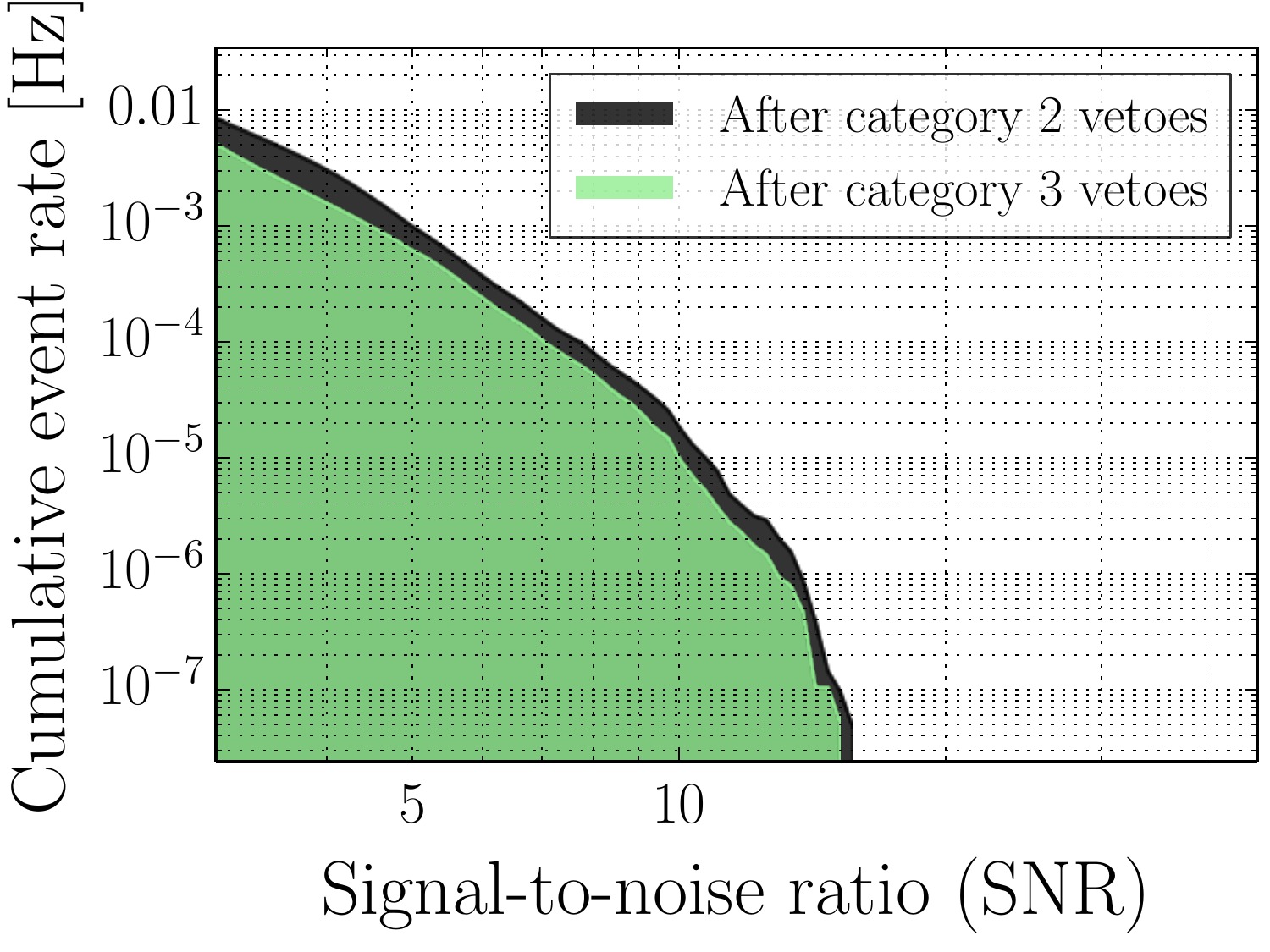}
        \includegraphics[width=.35\textwidth]{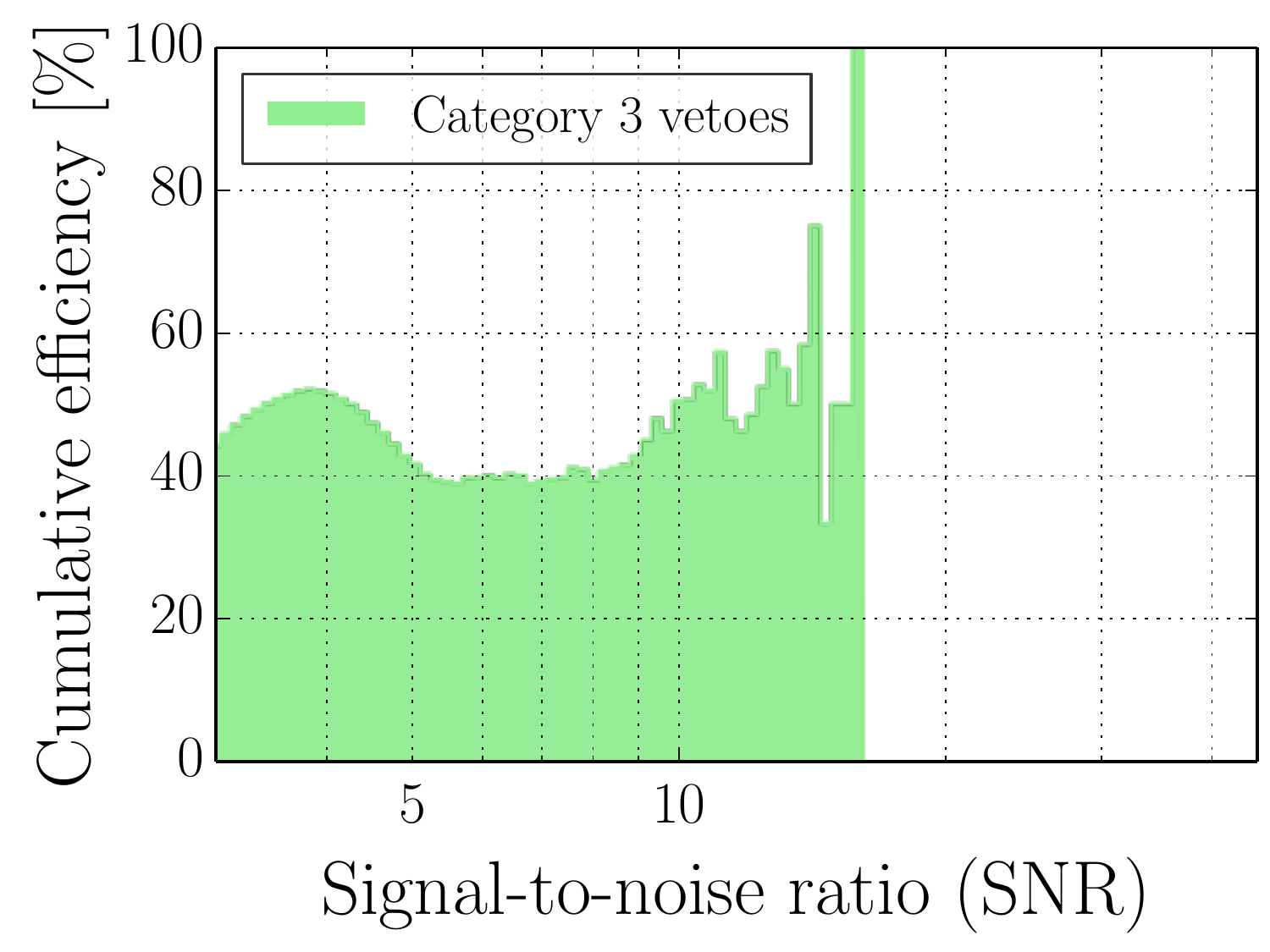}
        \label{subfig:h1_cwb_cat23}
    }\\
    \subfloat[L1]{%
        \includegraphics[width=.35\textwidth]{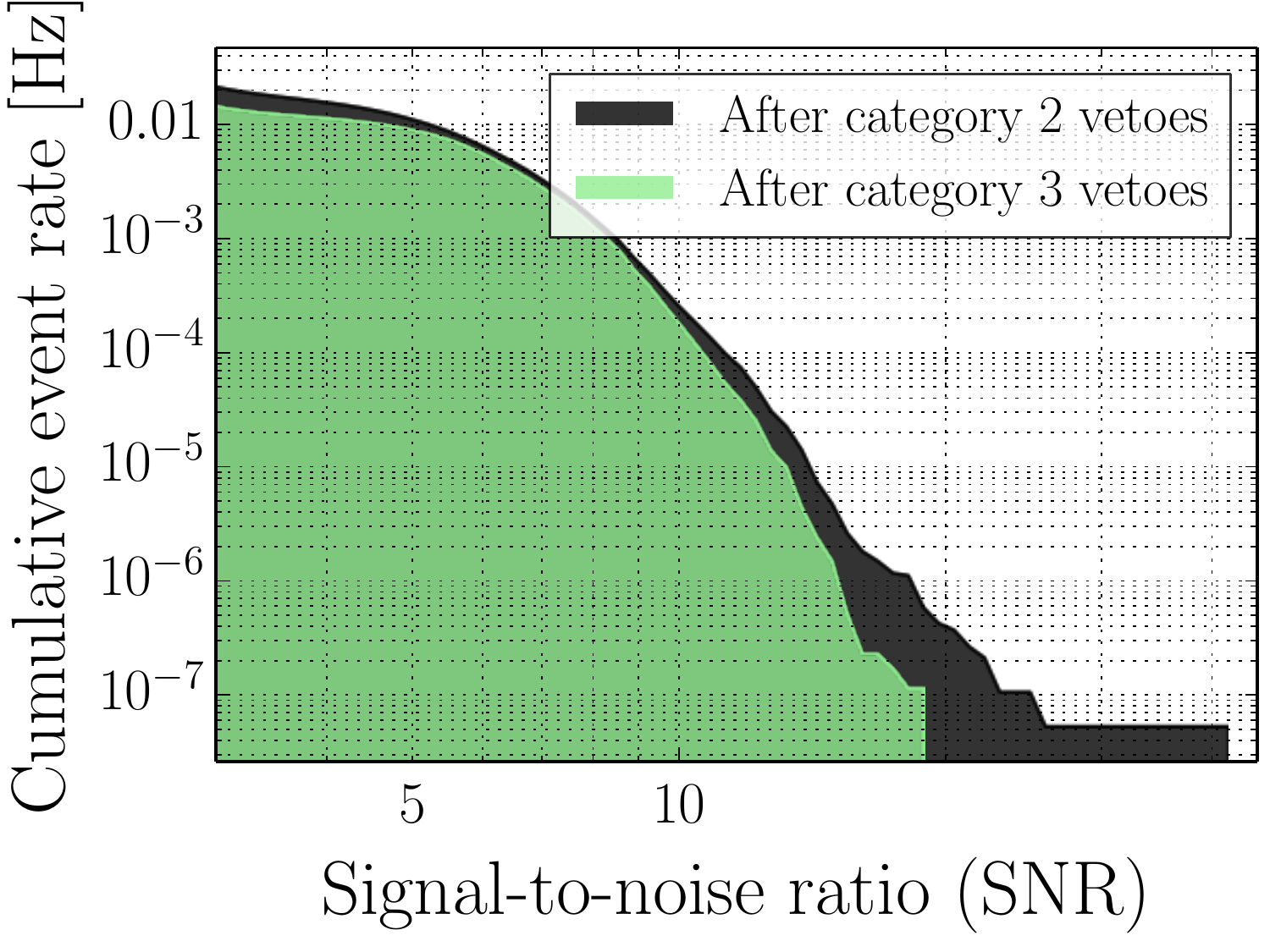}
        \includegraphics[width=.35\textwidth]{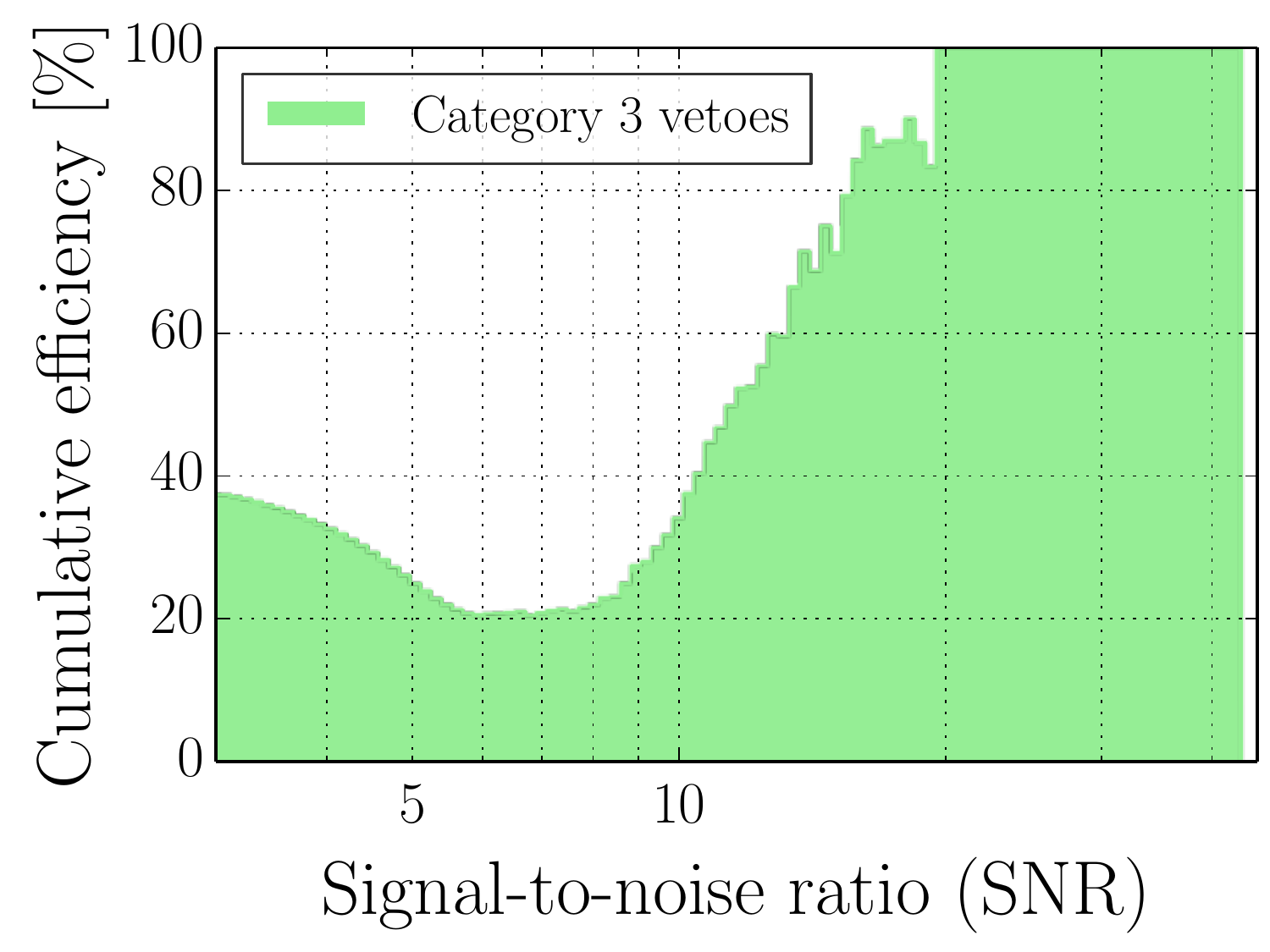}
        \label{subfig:l1_cwb_cat23}
    }
    \caption[The effect of category 3 vetoes on the \protect\ac{cwb} %
             pipeline]%
            {The effect of category 3 vetoes on the \protect\ac{cwb} %
             pipeline for \protect\subref{subfig:h1_cwb_cat23} H1 and %
             \protect\subref{subfig:l1_cwb_cat23} L1. %
             The left panels show the reduction in event rate, while the right %
             panels show the cumulative veto efficiency, both as a function of %
             single-detector \protect\ac{SNR}.}
    \label{fig:cwb_cat23}
\end{figure}

\Cref{fig:cwb_cat23} shows the effect of category 3 vetoes on the background events from the \ac{cwb} pipeline; these events were identified in the background from time time-slides and are plotted using the SNR reconstructed at each detector.
This search applies category 2 vetoes in memory, and does not record any events before this step, so efficiency statements are only available for category 3.
The results are shown after the application of a number of network- and signal-consistency checks internal to the pipeline that reject a large number of the loud events. As a result, the background is dominated by low \ac{SNR} events, with a small number of loud outliers.
At both sites, \ac{DQ} vetoes applied to this search have cumulative \ac{EDR} $\ge 5$ at \ac{SNR} 3, with those at L1 removing the tail above \ac{SNR} 20.
However, despite the reduction, this search was still severely limited by the remaining tail in the multi-detector background distribution~\cite{Abadie:2012rq}.

\begin{figure}[t]
    \centering
    \subfloat[H1]{%
        \includegraphics[width=.32\textwidth]{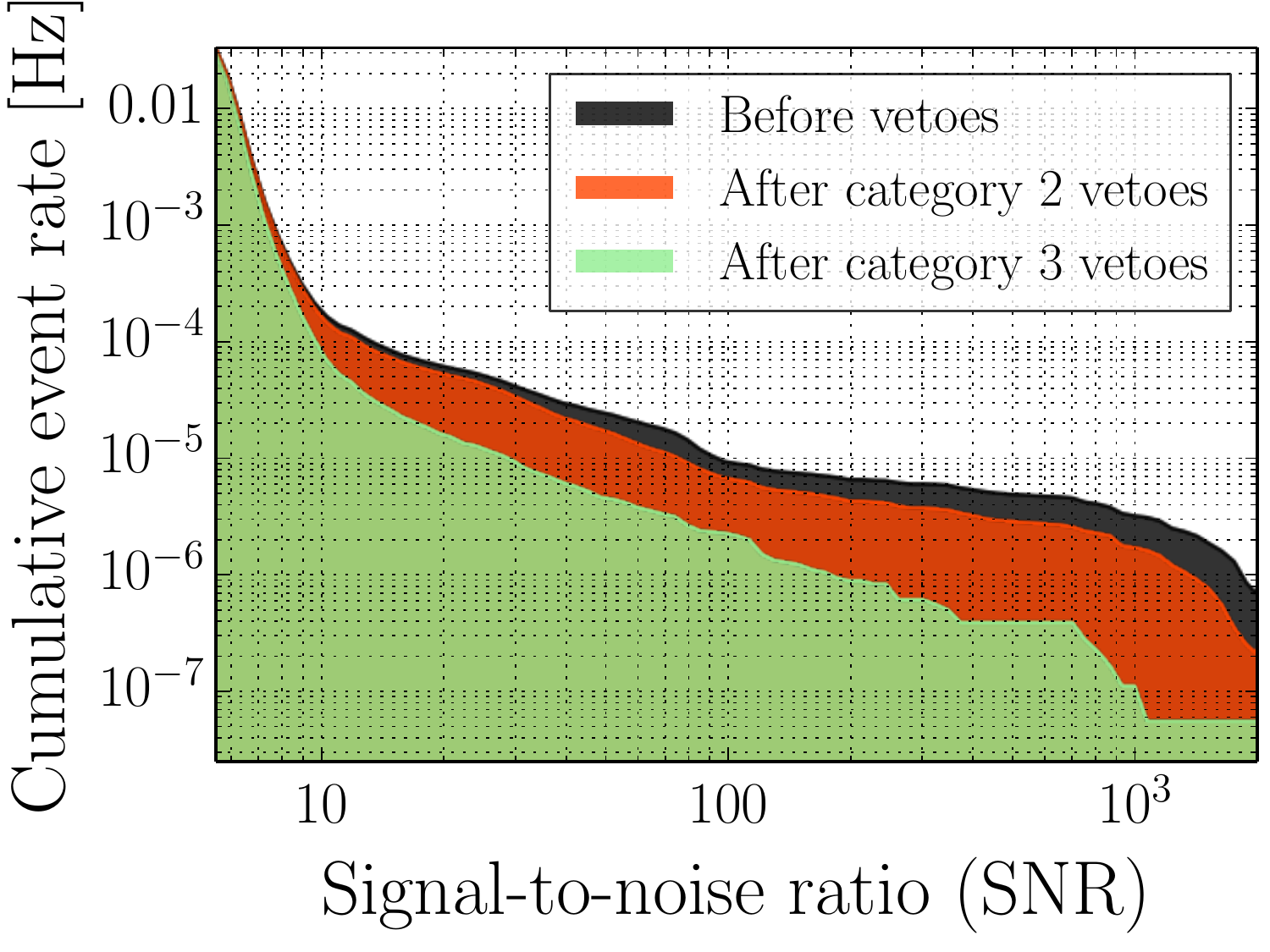}
        \includegraphics[width=.32\textwidth]{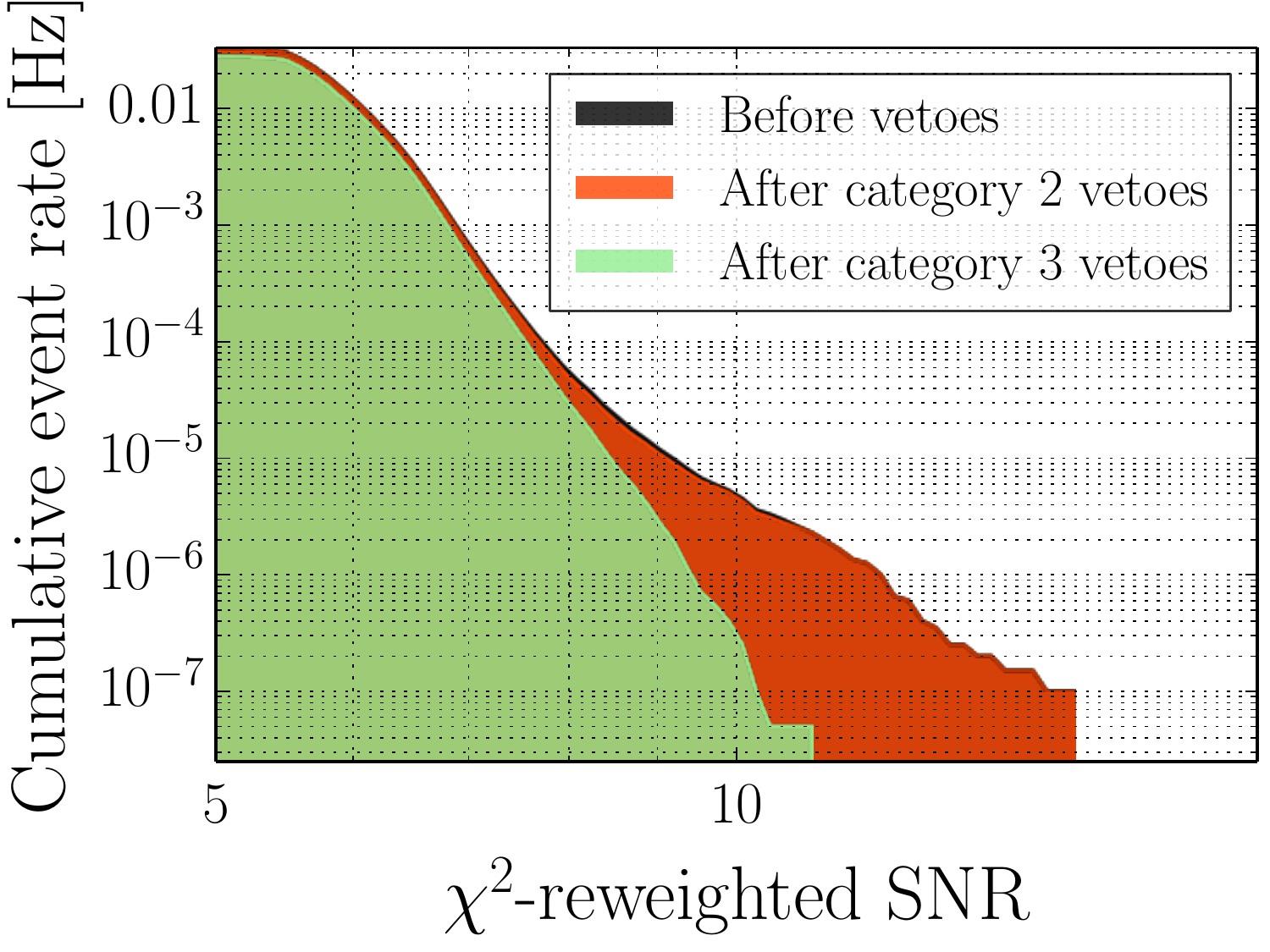}
        \includegraphics[width=.32\textwidth]{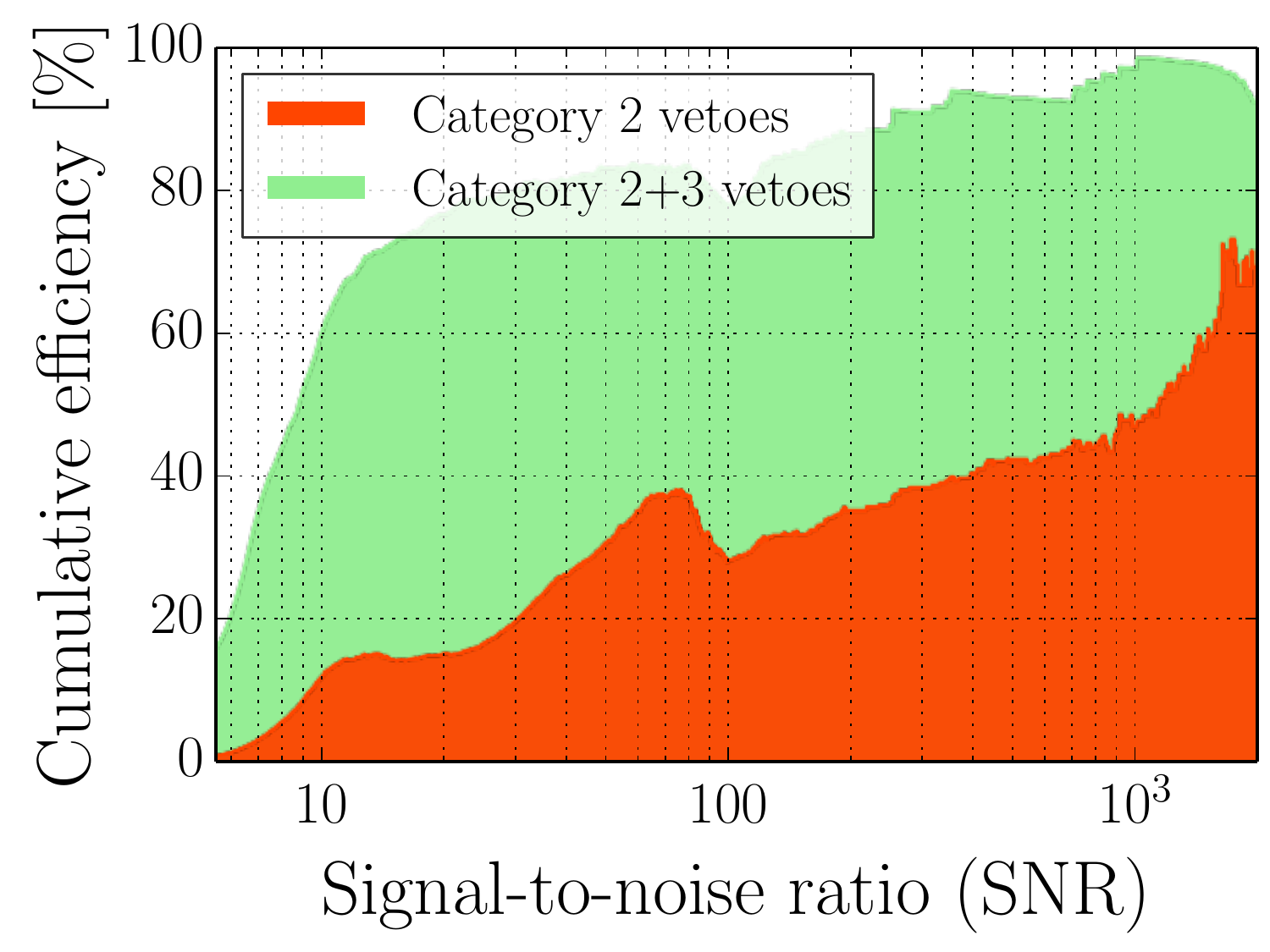}
        \label{subfig:h1_cbc_cat23}
    }\\
    \subfloat[L1]{%
        \includegraphics[width=.32\textwidth]{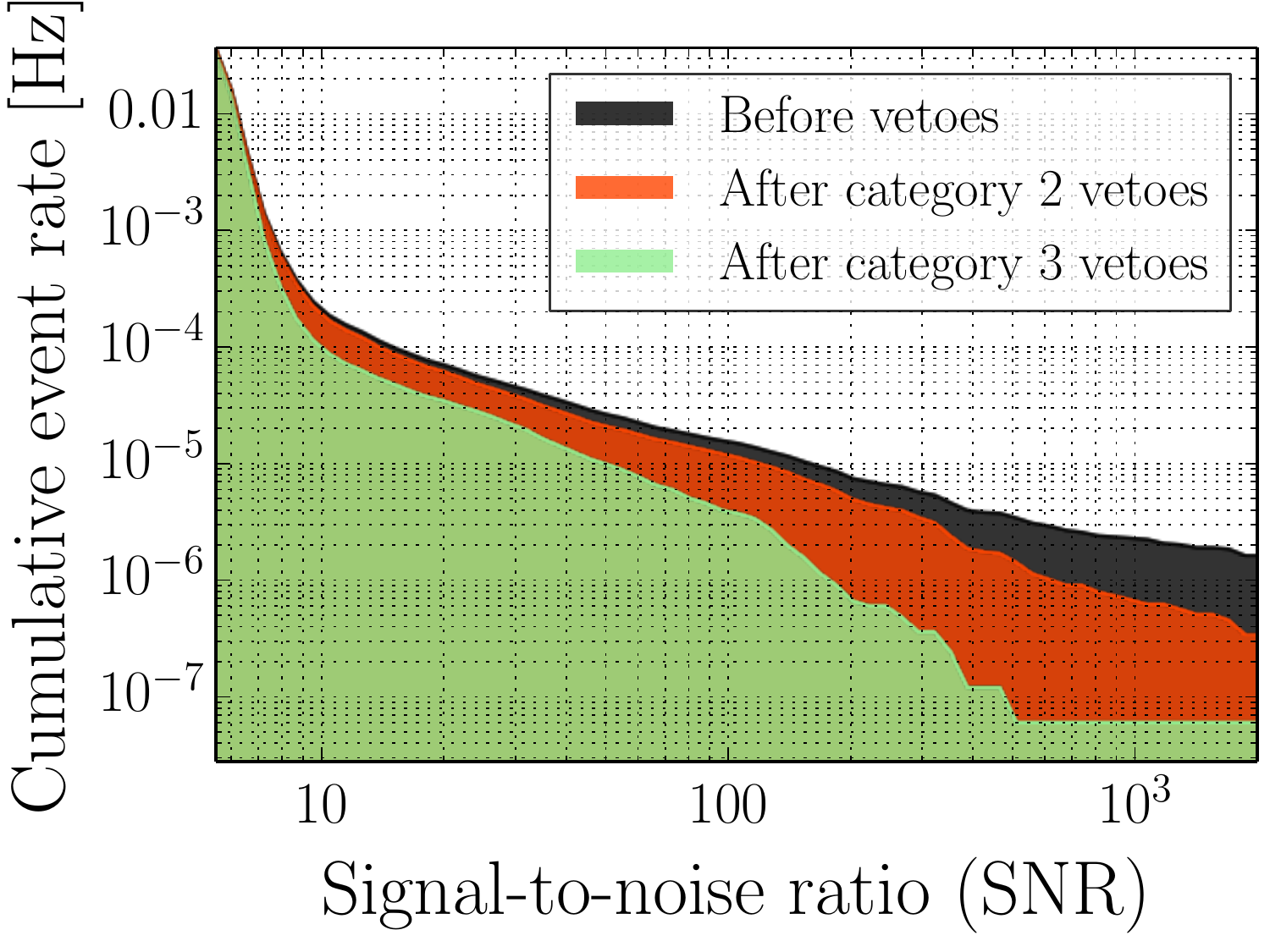}
        \includegraphics[width=.32\textwidth]{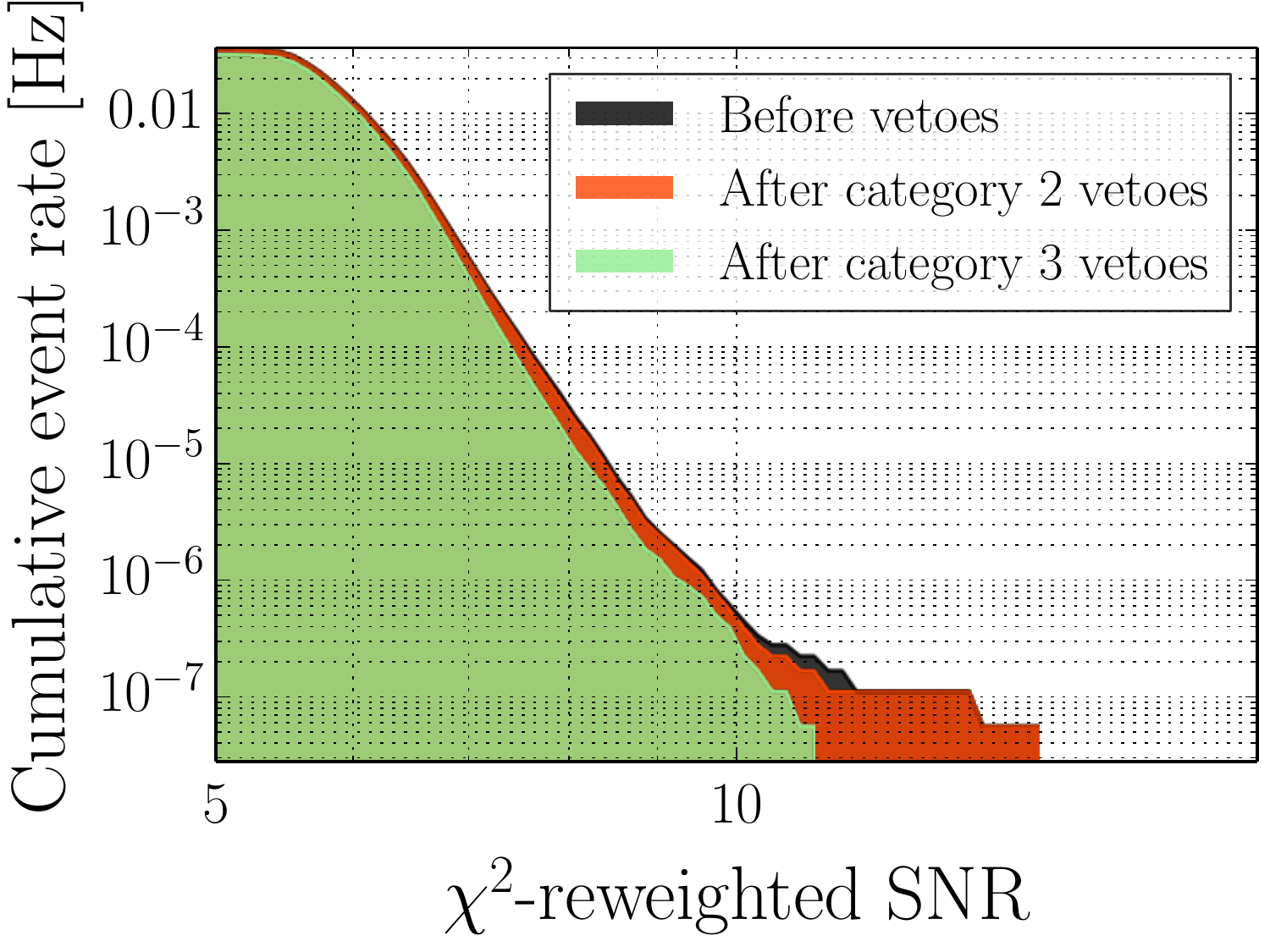}
        \includegraphics[width=.32\textwidth]{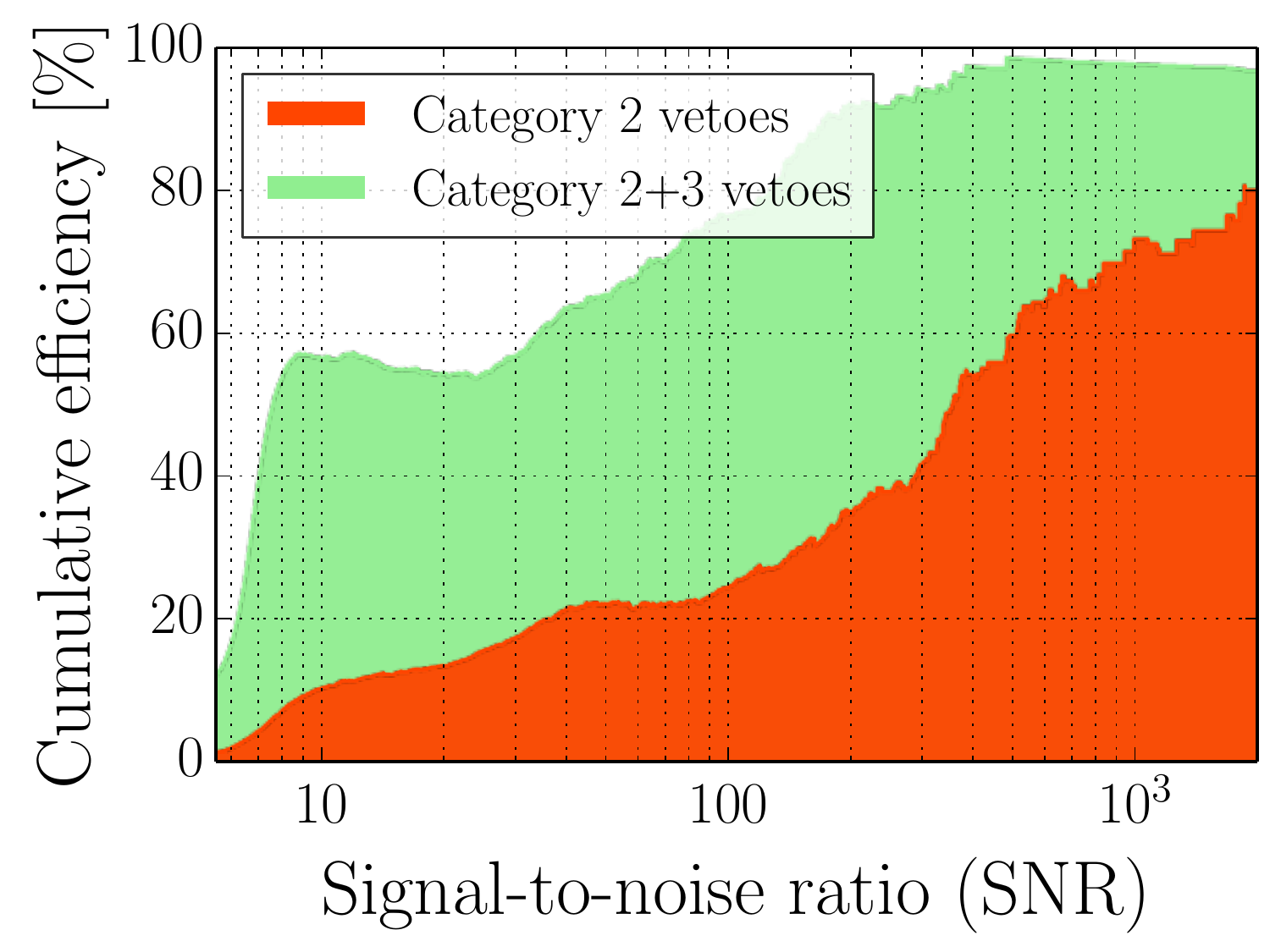}
        \label{subfig:l1_cbc_cat23}
    }
    \caption[The effect of category 2 and 3 vetoes on the CBC ihope pipeline]%
            {The effect of category 2 and jointly of category 2 and 3 vetoes %
             on the CBC ihope pipeline for %
             \protect\subref{subfig:h1_cbc_cat23} H1 and %
             \protect\subref{subfig:l1_cbc_cat23} L1. %
             The left panels show the reduction in event rate as a function of %
             \protect\ac{SNR}, the centre panels show the reduction in event %
             rate as a function of the $\chi^2$-weighted SNR, and the right %
             panels show the cumulative efficiency as a function of %
             \protect\ac{SNR}.}
    \label{fig:cbc_cat23}
\end{figure}

\Cref{fig:cbc_cat23} shows the effect of category 2 and 3 vetoes on the background from the \ac{CBC} ihope pipeline; this search sees a background extending to higher \ac{SNR}.
As shown, the background is highly suppressed by \ac{DQ} vetoes, with an efficiency of 50\% above \ac{SNR} 8, and 80\% above $\mathord{\sim} 100$ at both sites.
The re-weighted \ac{SNR} statistic, as defined in~\cite{Babak:2012zx}, is highly effective in down-ranking the majority of outliers with high matched-filter \ac{SNR}, but a non-Gaussian tail was still present at both sites.
Category 3 vetoes successfully removed this tail, reducing the loudest event at H1 (L1) from a re-weighted \ac{SNR} of 16.0 (15.3) to 11.1 (11.2).
Search sensitive distance was roughly inversely proportional to the $\chi^2$-weighted \ac{SNR} of the loudest event, and so reducing the loudest event by $\mathord{\sim}30$\% with $\mathord{\sim}10$\% deadtime can be estimated as a factor of $\mathord{\sim} 2.5$ increase in detectable event rate.


\subsection{Data quality in searches for long-duration signals}
\label{subsec:vetoes_cw}
In searches for both continuous \acp{GW} and a \ac{SGWB}, the duration and stationarity of data from each detector were the key factors in search sensitivity.
These analyses integrate over the entire science run in order to maximise the \ac{SNR} of a low-amplitude source.
Accordingly, they were impacted only very little by infrequent \glspl{glitch}, but were adversely affected by spectral lines and long periods of glitching in a given frequency band.

\subsubsection{Searches for continuous \protect\acsp{GW}.}
The PowerFlux pipeline~\cite{Dergachev:2005wr,Dergachev:2011wg} is one method used to conduct an all-sky search for \ac{GW} signals from pulsars.
This search, currently in progress, has chosen the final seven months of the \ac{S6} dataset in order to minimise the impact of poor detector performance from the earlier epochs.

A preliminary analysis of the data has shown instrumental features at high frequency causing the search sensitivity to drop towards that observed during \ac{S5}.
In all, $\mathord{\sim} 20$\% of frequency bands, each a few hundred mHz wide, have been identified as non-Gaussian, compared to almost zero in \ac{S5}.
Additionally, the noise due to beam jitter (see \cref{subsec:jitter}), has had a detrimental effect on sensitivity around 180-200\,Hz.

\subsubsection{Searches for a \protect\acs{SGWB}.}
For the \ac{S6} search for a \ac{SGWB}, \ac{DQ} cuts were made to eliminate data in H1 and L1 that were too noisy, too non-stationary, or that had apparent correlated noise between detectors%
\footnote{In the absence of a signal model, correlated noise and a \protect\ac{GW} signal are indistinguishable in a stochastic search.
However, a stochastic isotropic search assumes that the signal is broadband, and so narrow-band line features can be considered to be of instrumental, usually electronic, origin.}.
The analyses ran over times when both \ac{LIGO} detectors were taking science-quality data, excluding those times flagged as category 1 or category 4 veto segments.
The category 1 segments chosen for this search caused a 2\% reduction in coincident data for the \ac{LIGO} detector pair.

In addition, up to 5.5\% of data segments deviate from the stationary noise assumption, depending on frequency.
These were removed from the analysis by identifying segments whose standard deviation, $\sigma$, varies from neighbouring segments by greater than 20\%.
After applying all of the data quality cuts, $\mathord{\sim}$117~days of coincident live time for the \ac{LIGO} network remained.

Spectral noise lines are also a problem for the \ac{SGWB} search.
It is improbable to have a spectral noise line present in the same frequency bin (0.25\,Hz) in both H1 and L1, but it is possible.
In addition, a loud line in one detector can couple with a noise fluctuation in the other and produce an excess when the correlation is calculated between the two data streams.
In order to examine frequency bins for contamination, the coherence between two interferometers was calculated,
\begin{equation}
    \Gamma(f) = \frac{|\langle P_{12}(f) \rangle|^2}%
                     {\langle P_1(f) \rangle \langle P_2(f) \rangle},
\end{equation}
where $\langle P_{12} (f)\rangle$ is the average cross-spectral density and $\langle P_i(f) \rangle$ is the power spectral density for the $i^{\rm th}$ interferometer.
This was used to identify high coherence bins, searching at resolutions of 1\,Hz and 100\,mHz, using the method in \cite{Abbott:2009ws}.
This identified power line harmonics, 16~Hz harmonics from data acquisition, violin modes of the interferometer mirror suspension, and injected calibration signals.
These frequencies were excluded from the analysis, as were some frequency bins where a clear association with an environmentally produced noise line in either the H1 or L1 data could be made.
In total, 87 frequency bins (each 0.25\,Hz wide, in the range from 40--1000\,Hz) were removed from the \ac{S6} \ac{LIGO} \ac{SGWB} search.
The study of the coherence also revealed a small amount (0.2\%) of additional non-stationary time series data, and these were excluded.

In addition, the \ac{SGWB} search pipeline was run over \ac{LIGO} data after a non-physical time-shift had been applied.
The inspection of these data revealed further frequency bins where the \acl{SNR} was greater than $4.25$.
If frequency bins met this condition for at least 2 of the time shifted runs, they were removed from the final foreground analysis.
This removed 7 more frequency bins.\\

\noindent Preliminary results from the \ac{S6} \ac{CW} and \ac{SGWB} searches indicate that these steps have cleaned the data set, allowing more sensitive searches.
However, the increased non-stationarity and noise lines during \ac{S6} relative to \ac{S5} have produced a further detrimental effect on the data.
The \ac{S6} \ac{CW} searches can be expected to set better upper limits on \ac{GW} amplitudes than the \ac{S5} searches, nevertheless, spectral lines will appear as potential sources for all-sky \ac{CW} signal searches, and much work remains to explain the source of these presumed noise lines.
On the \ac{SGWB} side, the \ac{S6} data will provide a better upper limit as compared to the \ac{S5} results~\cite{Abbott:2009ws,Abadie:2011fx}.

It should also be noted that correlated magnetic field noise, from the Schumann resonances, was observed in correlations between magnetometers at H1, L1 and Virgo.
However it was determined that the level of correlated noise did not effect the S5 or S6 stochastic searches~\cite{Thrane:2013npa}.

    \section{Conclusions and outlook for Advanced LIGO}
\label{sec:conclusion}
The \ac{LIGO} instruments, at both Hanford and Livingston, are regularly affected by both non-Gaussian noise transients and long-duration spectral features.
Throughout \ac{S6} a number of problems were identified as detrimental to stable and sensitive data-taking at the observatories, as well as to the astrophysical searches performed on the data.

Instrumental fixes employed throughout the science run resulted in increasingly stable and sensitive instruments.
Median segment duration and overall duty factor improved from epoch to epoch (\cref{tab:segsummary}) and the detection range to the canonical binary neutron star inspiral increased by a significant factor (\cref{fig:range}).
\Acl{DQ} flags, used to identify known correlations between noise in auxiliary systems and the \ac{GW} data, allowed for a significant reduction in the event background of both core transient searches, ihope and \ac{cwb} (\cref{fig:cwb_cat23,fig:cbc_cat23}).
An \acl{EDR} above 5 for both searches, at both sites, allowed for a significant increase in the sensitivity of the search, improving the upper limits on event rate for both \ac{CBC} and generic \ac{GW} burst sources.

However, a tail of high \ac{SNR} events was still present in the \ac{cwb} search for \ac{GW} bursts, requiring deeper study of the glitch morphology and improved identification methods.
Additionally, the presence of noise lines outside the instrumental design had a detrimental, but not debilitating, effect on searches for long-duration signals.
A large number of these remaining transient and long-duration noise sources are still undiagnosed, meaning a large effort must be undertaken to mitigate similar effects in the second-generation instruments.

The first-generation \ac{LIGO} instruments were decommissioned shortly following the end of the science run (although immediately after \ac{S6} shot noise reduction was demonstrated in the H1 interferometer by using squeezed states of light~\cite{TheLIGOScientific:2013nha}), and installation and early testing of \ac{aLIGO} systems is now under way~\cite{Harry:2010zz}.
With the next data-taking run scheduled for 2015~\cite{LIGO:2013ip}, many methods and tools developed during the last run are set to be upgraded to further improve instrument and data quality.
Improvements are in place for each of the noise event detection algorithms, allowing for more accurate detection of transient noise in all channels, and work is ongoing for the \ac{hveto} and \ac{UPV} statistical veto generators~\cite{Essick:2013vga} to enable more efficient identification of sources of noise in the \ac{GW} data.
In addition, multi-variate statistical classifiers are being developed for use in glitch identification~\cite{Biswas:2013wfa}, using more information produced from event triggers to improve veto efficiency and identification of false alarms with minimal deadtime.

One of the major goals of the \ac{aLIGO} project is to contribute to multi-messenger astronomy -- the collaboration between \ac{GW} observatories and \ac{EM} and neutrino observatories~\cite{Abbott:2011ys,Evans:2012hd}.
Both the burst and \ac{CBC} search working groups are developing low-latency analyses from which to trigger followup with partner \ac{EM} telescopes, requiring a much greater effort in low-latency characterisation of the data.
With this in mind, a large part of the development in detector characterisation in the \ac{LSC} is now being devoted to real-time characterisation of instrumental data, including the \ac{GW} output and all auxiliary channels.
An \acl{ODC} system is being developed for \ac{aLIGO} that summarises the status of all instrumental and environmental systems in real-time to allow fast identification of false alarms in these on-line analyses, and reduce the latency of \ac{EM} follow-up requests.

Best estimates predict $\mathord{\sim}40$ detections of \acp{GW} from binary neutron star mergers per year at design sensitivity~\cite{Abadie:2010cf}, assuming stationary, Gaussian noise.
A great effort will be required in commissioning the new instruments to achieve these goals, including detailed characterisation of their performance before the start of the first advanced observing run.

\section{Acknowledgements}
The authors gratefully acknowledge the support of the United States
National Science Foundation for the construction and operation of the
LIGO Laboratory, the Science and Technology Facilities Council of the
United Kingdom, the Max-Planck-Society, and the State of
Niedersachsen/Germany for support of the construction and operation of
the GEO600 detector, and the Italian Istituto Nazionale di Fisica
Nucleare and the French Centre National de la Recherche Scientifique
for the construction and operation of the Virgo detector. The authors
also gratefully acknowledge the support of the research by these
agencies and by the Australian Research Council, 
the International Science Linkages program of the Commonwealth of Australia,
the Council of Scientific and Industrial Research of India, 
the Istituto Nazionale di Fisica Nucleare of Italy, 
the Spanish Ministerio de Econom\'ia y Competitividad,
the Conselleria d'Economia Hisenda i Innovaci\'o of the
Govern de les Illes Balears, the Foundation for Fundamental Research
on Matter supported by the Netherlands Organisation for Scientific Research, 
the Polish Ministry of Science and Higher Education, the FOCUS
Programme of Foundation for Polish Science,
the Royal Society, the Scottish Funding Council, the
Scottish Universities Physics Alliance, The National Aeronautics and
Space Administration, 
the National Research Foundation of Korea,
Industry Canada and the Province of Ontario through the Ministry of Economic Development and Innovation, 
the National Science and Engineering Research Council Canada,
the Carnegie Trust, the Leverhulme Trust, the
David and Lucile Packard Foundation, the Research Corporation, and
the Alfred P. Sloan Foundation.

    \vspace{5mm}
    \bibliographystyle{unsrt}
    \bibliography{references}
\end{document}